\pdfoutput=1
\documentclass[acmsmall,screen]{acmart}
\usepackage{placeins}
\usepackage{needspace}
\usepackage{array}
\usepackage{multirow}
\usepackage{colortbl}
\definecolor{rankhue}{HTML}{0072B2}
\newcommand{\rankfirst}[1]{\cellcolor{rankhue!32}\textbf{#1}}
\newcommand{\ranksecond}[1]{\cellcolor{rankhue!18}#1}
\newcommand{\rankthird}[1]{\cellcolor{rankhue!10}#1}
\usepackage{refcount}
\newcommand{\figrefcheck}[3]{\begingroup
  \edef\FigRefNow{\getrefnumber{#2}}\def\FigRefFig{#3}\def\FigRefNone{0}%
  \ifx\FigRefNow\FigRefFig\else\ifx\FigRefNow\FigRefNone\else
    \typeout{Figure numbers stale: figures/#1.pdf shows #3 for #2, text has \FigRefNow; rerun figures/#1_gen.py}%
  \fi\fi\endgroup}
\usepackage{tikz}
\usetikzlibrary{positioning,arrows.meta,decorations.pathreplacing,calc}
\acmJournal{TOIS}
\setcopyright{acmlicensed}
\acmDOI{XXXXXXX.XXXXXXX}
\acmYear{XXXX}
\copyrightyear{XXXX}
\acmMonth{0}
\acmVolume{X}
\acmNumber{X}
\acmArticle{X}


\begin{abstract}
Click-through rate (CTR) prediction underpins recommender systems and online advertising, where high-di\-men\-sion\-al categorical features interact in combinatorially many ways. Several recent CTR models, such as FinalNet and QNN, report higher accuracy when a network also learns from its own averaged predictions, for example by running the same network under different dropout patterns and training each run toward the average. Why this self-distillation helps, and whether its random sampling is needed, remains unclear. This work shows that as the number of dropout patterns grows without bound, the objective reduces exactly to the label loss of the logit-averaged prediction plus a penalty on how much the dropout copies of the network disagree. When dropout does not affect how a network gates its features, this penalty can be approximated in closed form, and FLMNet is designed on this principle and trains without sampling. Networks trained from different seeds disagree in the same form, and distilling several of them into one network recovers part of their ensemble gain. On five benchmarks against seventeen baselines, the resulting models take the top two places in accuracy with a single network at inference, and the gains of the penalty and of distillation, where measurable, largely add up.
\end{abstract}

\title{FLMNet: Self-Ensemble Distillation from Masks to Seeds
  for Single-Tower CTR Prediction}

\author{Yifan Wang}
\orcid{0009-0004-8721-4817}
\authornote{These authors contributed equally to this work and share first authorship.}
\authornote{Corresponding author: Yifan Wang.}
\affiliation{
  \department{Department of Mathematics}
  \institution{National University of Singapore}
  \city{Singapore}
  \country{Singapore}
}
\email{e1127286@u.nus.edu}

\author{Yifei Du}
\orcid{0009-0008-9872-9006}
\authornotemark[1]
\affiliation{
  \department{Hefei National Research Center for Physical Sciences at the Microscale and School of Physical Sciences; Shanghai Research Center for Quantum Science and CAS Center for Excellence in Quantum Information and Quantum Physics; Hefei National Laboratory}
  \institution{University of Science and Technology of China}
  \city{Hefei and Shanghai}
  \country{China}
}
\email{yifeidu@mail.ustc.edu.cn}

\renewcommand{\shorttitle}{FLMNet: Self-Ensemble Distillation from Masks to Seeds}
\AtBeginEnvironment{thebibliography}{\interlinepenalty=10000\relax}
\begin{document}
\maketitle

\section{Introduction}

Click-through rate (CTR) prediction estimates the probability that a user clicks a candidate item given user, item, and context features, and it drives ranking in recommender systems and online advertising. Production systems make predictions billions of times per day and replicate the trained model across data centers for serving~\cite{mcmahan2013}, so the deployed model is expected to answer each request with a single forward pass. The inputs are dozens of categorical fields, some with millions of distinct values, and whether a user clicks depends on how the values of different fields combine. The number of possible combinations grows multiplicatively with the number of fields and far exceeds what any training set covers, so a model cannot estimate each combination separately and must share parameters across them. Factorization machines share them by mapping each value to a low-dimensional embedding and scoring a combination by the inner product of its embeddings, so that every combination draws on all records in which its values occur~\cite{rendle2010}. Embeddings and the interactions computed from them have since been the starting point of neural CTR models.

Neural CTR models share a common template: each field value is mapped to an embedding, an interaction network combines the embeddings, and a linear head produces the logit. Successive designs have enriched the interaction network, from DeepFM, which feeds a shared embedding to both a factorization machine and a deep network~\cite{deepfm2017}, to explicit high-order modules such as the compressed interaction network of xDeepFM~\cite{xdeepfm2018} and the cross network of DCNv2~\cite{dcnv22021}. A recurring building block in this line is the Hadamard product, which raises the interaction order by multiplying a representation elementwise with a learned projection of the input. FINAL stacks Hadamard products into a factorized interaction layer, the basis of FinalNet~\cite{final2023}. QNN treats networks built on this operation, including factorization machines, cross networks, and FINAL, as neuron formats of a quadratic neural network, compares 25 of them under one setting, and builds its model, QNN-$\alpha$, on the strongest~\cite{qnn2025}. Both FinalNet and QNN report strong accuracy on public benchmarks.

Both models also add to the label loss a training signal that the model generates for itself. FinalNet computes predictions from two interaction blocks, averages their logits into a single score, and trains each block toward this aggregated score; the authors call this self-knowledge distillation~\cite{final2023}. QNN instead runs one network twice under independent dropout masks, averages the two predicted probabilities into a frozen target, and trains both passes toward this target; the authors call this a self-ensemble loss~\cite{qnn2025}. In both cases the teacher is an average over members of the model being trained, the members differ only in which block or which mask produces them, and no second model is trained or served. We refer to this family of objectives as self-ensemble distillation.

Both papers support the added term with ablations, and removing it lowers AUC on every dataset they report. The drop is smallest on Criteo for QNN, 0.06 percentage points, where its authors note that gains between well-tuned models are hard to obtain on large sparse data, and clearer on the smaller Frappe and KKBox datasets, 0.17 and 0.19 points~\cite{qnn2025}; for FinalNet it ranges from 0.03 points on Avazu to 0.12 on Frappe~\cite{final2023}. The two papers explain the gain differently, FinalNet as knowledge shared between blocks and QNN as mutual correction between passes that smooths the decision boundary, but neither runs an experiment that isolates which property of the objective produces it. What the self-ensemble objective regularizes, how its teacher should be formed, and whether the same effect can be obtained without extra stochastic forward passes therefore remain open, and these questions are the starting point of this paper.

\begin{figure}[htbp]
\centering
\includegraphics[width=\linewidth]{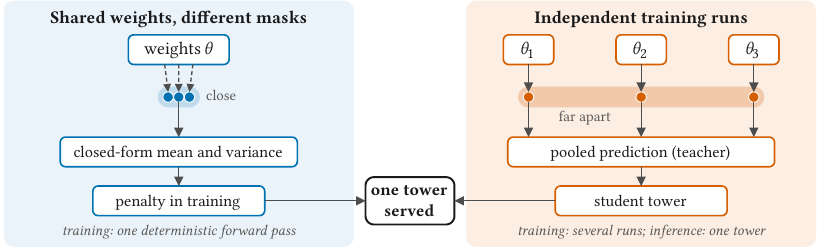}
\caption{Two kinds of ensemble members and their paths into a single deployed tower. Left: one set of weights under different masks; the members are close in prediction space, and their mean and variance enter training in closed form. Right: independently trained solutions; the members are far apart, and their pooled prediction serves as a teacher for a single student tower. In both cases only one tower is served.}
\Description{Conceptual comparison of shared-weight masking members and independently trained members, both feeding a single served tower.}
\label{fig:intro-members}
\end{figure}

These objectives and a classical ensemble average over different kinds of members. In FinalNet and QNN, the members belong to one model and differ only in the block or dropout mask that produces the prediction, so the teacher is an average taken at a single set of weights. In a classical ensemble, the members are independently trained networks, and the teacher averages solutions reached from different initializations~\cite{hinton2015} or data orders. The two averages are not interchangeable: functions sampled from the dropout neighborhood of one solution stay within a single mode in prediction space, whereas independently trained networks explore different modes~\cite{fort2019}. Pooled in logit space, however, both averages induce a gap of the same form, the mean of the softplus minus the softplus of the mean, because the label term of the cross-entropy loss is linear in the logit; the gap is taken over masks in one case and over solutions in the other. Distilling an ensemble of independently trained models into one network is an established route to a stronger single CTR model~\cite{ensemblekd2020}, with teachers that differ in architecture or in the split of the training data. Our experiments on Criteo suggest that the gain requires neither difference (Section~\ref{sec:exp-seed}), in line with the finding in speech recognition that random initialization alone gives sufficient diversity~\cite{hinton2015}. Towers of one architecture that differ only in their seed pass more than half of the gain of a five-tower ensemble to a student, and towers that differ only in the order of the training examples disagree almost as much as towers from different seeds (Appendix~\ref{app:order-only}). The student starts from a converged tower, where the training labels no longer point toward parameters that generalize better but the pooled logit of the other seeds still does. Figure~\ref{fig:intro-members} sketches both kinds of member and how each reaches a single deployed tower.

Codistillation shows that distillation between networks of one architecture can improve CTR models. Two copies of a network are trained together, each distilling from the other, and on Criteo either copy matches the log loss of their two-model ensemble~\cite{anil2018}. The authors attribute the gain mainly to the copies seeing different subsets of the data and leave open why a teacher's mistakes help the student.

In CTR prediction, FinalNet and QNN are two of several models that add auxiliary losses built from their own sub-networks or from perturbed views of their features~\cite{fsdnet2024,kdef2025,fcn2026,cetn2025,cl4ctr2023}. Their authors explain the gains differently: FSDNet credits the guidance its deepest layer gives the shallower ones~\cite{fsdnet2024}, FCN credits a label loss tailored to each of its sub-networks~\cite{fcn2026}, and CL4CTR credits better representations of rare features learned through contrastive losses~\cite{cl4ctr2023}. None of these papers, FinalNet and QNN included, identifies what the added loss regularizes. Outside CTR prediction, dropout, its consistency variants, and distillation among networks of one architecture have been studied for other tasks, from linear regression to language modeling, image classification, and sequential recommendation~\cite{fastdropout2013,wager2013,baldi2013,dropout2014,rdrop2021,wei2020,allenzhu2020,dml2018,one2018,kdcl2020,okddip2020,snapshot2019,emkd2023,csrec2025,flame2026}, but their explanations do not directly account for the gains in CTR models. For members trained independently, transferring the gain into a single model is established; what remains open is whether the gains from masked members and from independently trained members are distinct, and whether one tower can collect both. Prompted by these methods, we trace where their gain comes from, with ablations that vary the training objective (Section~\ref{sec:exp-mask}), the seed (Section~\ref{sec:exp-seed}), and the data order (Appendix~\ref{app:order-only}).

This paper asks three questions. What does the masked self-ensemble objective become as the number of masks grows without bound, and what does the limit regularize? Which tower structure gives the moments of the logit over masks in closed form, so that the limit can be trained with one deterministic forward pass? And can the further gain of independently trained solutions be collected in the same tower, and at what training cost?

Our answer to the first two questions is analytic. For members pooled in logit space, as the number of masks grows without bound, the masked self-ensemble objective converges to the label loss at the mean logit plus a multiple of the Jensen gap of the softplus over masks, up to a term with zero gradient. The limit therefore regularizes the disagreement among masked members and nothing else. Ordinary dropout training is the special case with multiplier one. In the limit, the QNN objective with its pool taken in logit space has multiplier two, so only the excess over one is distillation added beyond dropout. Training replaces the gap by a Gaussian surrogate, which needs only the mean and variance of the logit over masks. Both moments have closed forms under three conditions: the multiplicative gates of the tower read only unmasked inputs, the readout is affine, and each layer rescales its running representation coordinatewise. The mean logit then equals the logit of the forward pass without dropout, and an inexpensive approximation of the exact variance lets the surrogate be evaluated within one deterministic forward pass.

QNN-$\alpha$, the strongest baseline on average in our experiments, does not meet the first condition: its gate reads the masked output of the previous layer, so a closed form misstates the mean and variance of its logit over masks. Letting its gate read the input embedding makes the closed form exact, but on Criteo this change lowers validation AUC by $10.5\times10^{-4}$ even without the penalty. We therefore design FLMNet, a single tower whose field-local gate units each read one field of the unmasked embedding, with a learned mixer that spreads each field's gate over the other fields. FLMNet satisfies all three conditions by construction, and without the penalty its validation AUC comes within $0.9\times10^{-4}$ of that of the original QNN-$\alpha$.

The third question turns from masks to seeds. Towers trained from different seeds also disagree, and pooling their logits lowers Logloss below that of the average member by exactly the gap across seeds. This gap has no closed form, but part of the gain can be moved into one tower by distillation: FLMNet-KD starts from one FLMNet-CF tower, the tower trained with the closed-form penalty, and is trained to match the pooled logit of $K$ such towers, at the cost of $K+1$ training runs. We call the part of a member's logit that departs from the pooled logit its noise: mask noise across the masks of one run, and seed noise across seeds. The pooled logit keeps what the members share, and the noise is what each member fits on its own; in our experiments, suppressing it mainly holds back overfitting (Sections~\ref{sec:exp-mask} and~\ref{sec:exp-seed}). The penalty suppresses mask noise and distillation suppresses seed noise, and the deployed model is in both cases a single tower that predicts in one forward pass.

Our contributions are as follows.
\begin{itemize}
\item We derive the infinite-masking limit of self-ensemble distillation for members pooled in logit space, show that it regularizes only the Jensen gap among masked members, with ordinary dropout as the special case of multiplier one, and bound the gap above and below in terms of the variance of the logit.
\item We isolate three conditions under which the mean and variance of the logit over masks have closed forms, and design FLMNet, a single tower that satisfies them and trains the surrogate of the limit without sampled forward passes, with a step time close to that of QNN-$\alpha$.
\item On five public benchmarks against seventeen baselines, FLMNet-CF and FLMNet-KD take the top two places in both AUC and Logloss, and on the benchmarks where each gain alone exceeds the variation across seeds, the gains of the closed-form penalty and of cross-seed distillation largely add up.
\end{itemize}

\section{Preliminaries}\label{sec:prelim}

A CTR instance consists of $C$ categorical fields and a binary label $y\in\{0,1\}$. Each field value is mapped to an embedding of dimension $d$, the $C$ embeddings are concatenated into $X\in\mathbb{R}^{Cd}$, and a tower $f_\theta$ maps $X$ to a logit $z=f_\theta(X)$; the predicted click probability is $p=\sigma(z)$ with $\sigma$ the sigmoid. Training minimizes the binary cross-entropy $\mathrm{CE}(y,\sigma(z))=-y\log\sigma(z)-(1-y)\log(1-\sigma(z))$, and we use the same notation for a soft target $t\in[0,1]$. Writing $\varsigma(z)=\log(1+e^{z})$ for the softplus, so that $\log\sigma(z)=z-\varsigma(z)$ and $\log(1-\sigma(z))=-\varsigma(z)$, the cross-entropy becomes
\begin{equation}
\mathrm{CE}(t,\sigma(z))=\varsigma(z)-t\,z.
\label{eq:ce-softplus}
\end{equation}
Every cross-entropy in this paper is therefore a convex function of the logit minus a term linear in the logit, whether the target is a label or a soft prediction. The derivatives $\varsigma'=\sigma$ and $\varsigma''=\sigma(1-\sigma)$ will be used in Section~\ref{sec:limit}; \eqref{eq:ce-softplus} is the only property of the loss that the derivations there rely on.

Dropout perturbs the tower with a random mask $m$ drawn from a distribution $\Pi$: each maskable unit is multiplied by an independent factor $s$ that equals $0$ with probability $\rho$ and $1/(1-\rho)$ otherwise, so that $\mathbb{E}[s]=1$. This inverted scaling is the default in current frameworks and is assumed throughout; Section~\ref{sec:tower} relies on it. We write $z_m=f_\theta(X;m)$ for the logit under mask $m$, $\mu=\mathbb{E}_m[z_m]$ for its mean, $\delta=z_m-\mu$ for the perturbation, which we call the mask noise of the logit, and $V=\mathrm{Var}_m(z_m)$ for its variance. Because $\Pi$ has finite support, $\mathbb{E}_m[\cdot]$ is a finite sum; $S$ forward passes with independent masks provide a Monte Carlo estimate of it, and the limit $S\to\infty$ taken in Section~\ref{sec:limit} refers to this expectation, not to an infinite number of models. The estimate carries a sampling error, which for the pooled logit of $S$ masks has variance $V/S$; it is set by the mask noise but differs from it, since it vanishes as $S\to\infty$ while the mask noise remains.

Two kinds of members appear in this paper. Masked members share one parameter vector $\theta$ and differ only in the mask; their logits are $z_m$ and their pool is the mean logit $\mu$. Independent members are $K$ towers trained from different seeds $s=1,\dots,K$, with mean logits $\mu_s$ and pool $\bar\mu_K=\frac1K\sum_{s}\mu_s$ (Section~\ref{sec:seedgap}). In both cases members are pooled in logit space, and in both cases the target term of \eqref{eq:ce-softplus} is linear in the logit. Consequently every identity in Section~\ref{sec:limit} that is derived for masked members by taking $\mathbb{E}_m$ holds verbatim for independent members with $\mathbb{E}_m$ replaced by the average over $k$; the two kinds differ in what the average ranges over, not in the algebra.

\section{Self-Ensemble Distillation at the Infinite-Masking Limit}
\label{sec:limit}
\subsection{Finite-Mask Self-Ensemble Objective}
We start from the training objective of QNN-$\alpha$~\cite{qnn2025}, which its authors call the self-ensemble loss. During training the tower is run twice on the same input with independent masks $m_1,m_2$, giving probabilities $q_i=\sigma(z_{m_i})$ and their mean $\bar q=\tfrac12(q_1+q_2)$. In the notation of \eqref{eq:ce-softplus}, the released implementation minimizes
\begin{equation}
\mathcal L_{S=2}=\mathrm{CE}(y,\bar q)+\mathrm{CE}(\mathrm{sg}[\bar q],q_1)+\mathrm{CE}(\mathrm{sg}[\bar q],q_2),
\label{eq:se-loss}
\end{equation}
where $\mathrm{sg}[\cdot]$ denotes stop-gradient. Three features of this objective matter for what follows. The label term is applied to the ensemble $\bar q$, not to the individual members. The teacher is the arithmetic mean of member probabilities with its gradient stopped, so the last two terms distill the ensemble into each member; this is the sense in which we call the objective self-ensemble distillation. Finally, the distillation terms are summed over members rather than averaged, with no separate weight, which ties the strength of distillation to the number of passes; the next paragraph undoes this coupling.

Two changes make \eqref{eq:se-loss} suitable for letting $S$ grow. First, the sum of distillation terms diverges with $S$, so we give the objective $S$ members and a weight $\lambda\ge0$ on the distillation term,
\begin{equation}
\mathcal L_S=\mathrm{CE}(y,\mathcal E_S)+\lambda\,\frac1S\sum_{i=1}^{S}\mathrm{CE}\big(\mathrm{sg}[\mathcal E_S],q_i\big),\qquad q_i=\sigma(z_{m_i}),
\label{eq:normalized}
\end{equation}
which recovers \eqref{eq:se-loss} at $S=2$, $\lambda=2$. The weight and the number of passes are now independent axes, whereas the original form ties them together. Second, the teacher $\mathcal E_S$ in \eqref{eq:normalized} is a pooling rule still to be fixed; the authors take the arithmetic mean of the $q_i$. We let the distillation term select it. For a fixed target $t$, \eqref{eq:ce-softplus} gives $\mathrm{CE}(t,q_i)=H(t)+\mathrm{KL}(t\,\|\,q_i)$ with $H$ the binary entropy, so each member is pulled toward the teacher in the divergence $\mathrm{KL}(t\,\|\,\cdot)$. We take as teacher the point closest to all members in that same divergence.

\begin{lemma}
\label{lem:pool}
For fixed $q_i=\sigma(z_{m_i})$, $i=1,\dots,S$, the unique minimizer over $t\in(0,1)$ of $\sum_i\mathrm{KL}(t\,\|\,q_i)$ is
\begin{equation}
t^\ast=\sigma(\bar z_S),\qquad \bar z_S:=\frac1S\sum_{i=1}^{S}z_{m_i}.
\label{eq:teacher}
\end{equation}
\end{lemma}

\begin{proof}
By the identities preceding \eqref{eq:ce-softplus}, $\log q_i-\log(1-q_i)=z_{m_i}$, so the derivative in $t$ is $\sum_i\big[\log\tfrac{t}{1-t}-z_{m_i}\big]=S\big[\log\tfrac{t}{1-t}-\bar z_S\big]$, and the objective is strictly convex in $t$.
\end{proof}

Hence $\mathcal E_S:=\sigma(\bar z_S)$. This is the normalized geometric mean of the member distributions, also known as the logarithmic opinion pool~\cite{Heskes1997LOP}, and it depends on the members only through their mean logit; the arithmetic mean of probabilities is instead the minimizer of $\sum_i\mathrm{KL}(q_i\,\|\,t)$, the opposite direction from the one the distillation term uses. Since $\mathcal E_S$ also enters the label term, the label is then fitted by the same pooled prediction that serves as the teacher. Two consequences follow. As $S\to\infty$, $\bar z_S\to\mu$ and the teacher becomes $\sigma(\mu)$, so the label term reduces by \eqref{eq:ce-softplus} to $\varsigma(\mu)-y\mu$; and Section~\ref{sec:tower} builds the tower so that $\mu$ is the output of its mask-free forward pass, so the limiting teacher costs no additional computation. The block-wise objective of FINAL~\cite{final2023} has the same label-plus-distillation form and already pools its two blocks in logit space, but its members are fixed sub-networks rather than random draws, so $S$ is not an axis for it and the limit taken next does not apply.

\subsection{Infinite-Masking Limit}
\label{sec:limit-inf}

We now let $S\to\infty$. The mask distribution $\Pi$ has finite support and the loss is bounded on it, so the averages over the $S$ draws converge to their expectations, $\bar z_S\to\mu$, and \eqref{eq:normalized} becomes
\begin{equation}
\mathcal L_\infty=\mathrm{CE}\big(y,\sigma(\mu)\big)+\lambda\,\mathbb E_m\Big[\mathrm{CE}\big(\mathrm{sg}[\sigma(\mu)],\sigma(z_m)\big)\Big].
\label{eq:limit}
\end{equation}
This is a limit of the training objective, not of the training trajectory. At finite $S$ the objective is not an unbiased estimate of \eqref{eq:limit}: by Proposition~\ref{prop:finite}, its expected gradient corresponds to the multiplier $\kappa_S$ rather than $\kappa$, and it also carries sampling error. Both disappear in the limit; the mask noise that the limit penalizes does not. Wei et al. separate these two effects for dropout and report that its explicit part alone, the expected objective, does not by itself reach the accuracy of single-sample dropout~\cite{wei2020}. Whether the same holds here is an empirical question, and we measure it in Section~\ref{sec:exp}.

Two facts turn \eqref{eq:limit} into a closed expression. Write $A:=\mathbb E_m[\varsigma(z_m)]-\varsigma(\mu)$ for the Jensen gap of the softplus; it is nonnegative because $\varsigma$ is convex.

\begin{lemma}
\label{lem:ambiguity}
For any target $t$ that does not depend on $m$,
\begin{equation}
\mathbb E_m\big[\mathrm{CE}(t,\sigma(z_m))\big]=\mathrm{CE}\big(t,\sigma(\mu)\big)+A.
\label{eq:ambiguity}
\end{equation}
\end{lemma}

\begin{proof}
\begin{align*}
\mathbb E_m\big[\mathrm{CE}(t,\sigma(z_m))\big]
&\overset{\eqref{eq:ce-softplus}}{=}\mathbb E_m\big[\varsigma(z_m)\big]-t\,\mu\\
&=\varsigma(\mu)+A-t\,\mu\\
&\overset{\eqref{eq:ce-softplus}}{=}\mathrm{CE}\big(t,\sigma(\mu)\big)+A.
\end{align*}
\end{proof}

At $t=\sigma(\mu)$ the left side is the expected cross-entropy from the teacher to its members and the first term on the right is the teacher's entropy, so $\mathbb E_m[\mathrm{KL}(\sigma(\mu)\,\|\,\sigma(z_m))]=A$: the gap is the expected divergence from the pooled prediction to the members it pools. Lemma~\ref{lem:ambiguity} is the ambiguity decomposition of Krogh and Vedelsby~\cite{Krogh1994Ensembles}, stated there for squared loss, in the form it takes for logarithmic pooling~\cite{Heskes1997LOP}; the general Bregman statement is due to Wood et al.~\cite{Wood2023Diversity}, and concurrent work on CTR prediction writes it for the binary cross-entropy of a mean-logit ensemble~\cite{recap2026}. What is specific to this paper is the alignment: the gap the decomposition isolates is the distillation term of \eqref{eq:limit} itself. The argument uses only the convexity of $\varsigma$ and $\mathbb E_m[\delta]=0$, so it holds for any exponential family with $\varsigma$ replaced by the log-partition $\Phi$ and $A$ by $\mathbb E_m[\Phi(z_m)]-\Phi(\mu)$. For the Gaussian, $\Phi(\eta)=\eta^2/2$ and $A=\tfrac12\mathrm{Var}_m(z_m)$ exactly, which is the squared-loss statement cited above.

Nothing so far fixes where the label term is applied. Placing it on each member instead of on the pool gives
\begin{equation}
\mathcal L_S'=\frac1S\sum_{i=1}^{S}\mathrm{CE}(y,q_i)+\lambda\,\frac1S\sum_{i=1}^{S}\mathrm{CE}\big(\mathrm{sg}[\mathcal E_S],q_i\big),
\label{eq:member-label}
\end{equation}
the placement under which every member is trained against the label. Both forms are covered by the following statement.

\begin{theorem}
\label{thm:limit}
Write $\doteq$ for equality up to a term whose gradient vanishes. As $S\to\infty$, both \eqref{eq:normalized} and \eqref{eq:member-label} satisfy
\begin{equation}
\mathcal L_\infty\doteq\mathrm{CE}\big(y,\sigma(\mu)\big)+\kappa A,
\label{eq:main}
\end{equation}
with $\kappa=\lambda$ for \eqref{eq:normalized} and $\kappa=1+\lambda$ for \eqref{eq:member-label}.
\end{theorem}

\begin{proof}
Write $t=\mathrm{sg}[\sigma(\mu)]$, so that $\mathrm{CE}(t,\sigma(\mu))=H(t)$ and $H(t)$ carries no gradient. For \eqref{eq:normalized},
\begin{align*}
\mathcal L_\infty
&\overset{\eqref{eq:limit}}{=}\mathrm{CE}\big(y,\sigma(\mu)\big)+\lambda\,\mathbb E_m\big[\mathrm{CE}(t,\sigma(z_m))\big]\\
&\overset{\eqref{eq:ambiguity}}{=}\mathrm{CE}\big(y,\sigma(\mu)\big)+\lambda\big[H(t)+A\big]\\
&\doteq\mathrm{CE}\big(y,\sigma(\mu)\big)+\lambda A.
\end{align*}
For \eqref{eq:member-label}, the same limit applies to each of its two terms, and
\begin{align*}
\mathcal L_\infty'
&=\mathbb E_m\big[\mathrm{CE}(y,\sigma(z_m))\big]+\lambda\,\mathbb E_m\big[\mathrm{CE}(t,\sigma(z_m))\big]\\
&\overset{\eqref{eq:ambiguity}}{=}\big[\mathrm{CE}\big(y,\sigma(\mu)\big)+A\big]+\lambda\big[H(t)+A\big]\\
&=\mathrm{CE}\big(y,\sigma(\mu)\big)+(1+\lambda)A+\lambda H(t)\\
&\doteq\mathrm{CE}\big(y,\sigma(\mu)\big)+(1+\lambda)A.
\end{align*}
\end{proof}

Three consequences follow. First, setting $\lambda=0$ in \eqref{eq:member-label} leaves $\mathbb E_m[\mathrm{CE}(y,\sigma(z_m))]$, the expected objective of ordinary dropout training, for which Theorem~\ref{thm:limit} gives $\kappa=1$. The self-ensemble objective is not a different kind of objective but the same one with $\kappa$ pushed above one, so it is $\kappa-1$, not $\kappa$, that measures the distillation added beyond dropout; the original form \eqref{eq:se-loss}, with its pool taken in logit space, sits at $\kappa=2$ in the limit; at the $S=2$ used in training, its expected gradient corresponds to $\kappa_2\approx1.5$ (Proposition~\ref{prop:finite}). Second, $A\ge0$ gives $\mathcal L_\infty\ge\mathrm{CE}(y,\sigma(\mu))$, with equality only when all members agree. The objective cannot be driven below the loss of its own ensemble; what it buys is a single prediction that behaves like one. This concerns the training loss and carries no guarantee about test accuracy. Third, $A$ enters with a positive coefficient, so minimizing $\mathcal L_\infty$ drives the members toward the pooled prediction, that is, it suppresses the mask noise: the objective spends the ensemble's disagreement rather than preserving it. Intuitively, each mask gives a member a different partial view of the input. Pulling the members together therefore favors patterns that all views support over idiosyncrasies of the training data that only some views can fit, so the penalty should discourage memorization and slow overfitting; Section~\ref{sec:exp-mask} tests this prediction.

The limit also fixes what a finite $S$ does. Write $A(S):=\mathbb E\big[\varsigma(\bar z_S)\big]-\varsigma(\mu)$ for the gap of a pool of $S$ independent members, and $r_S:=A(S)/A$ for $A>0$.

\begin{proposition}
\label{prop:finite}
Fix $\rho$, so that the mask distribution does not depend on $\theta$. For every $S\ge1$, the expected gradient of \eqref{eq:normalized} over the $S$ masks is the gradient of $\mathrm{CE}(y,\sigma(\mu))+\lambda A+(1-\lambda)A(S)$, and that of \eqref{eq:member-label} is the gradient of $\mathrm{CE}(y,\sigma(\mu))+(1+\lambda)A-\lambda A(S)$. At each $(\theta,X)$ the two penalties equal $\kappa_SA$ with $\kappa_S=\lambda+(1-\lambda)r_S$ and $\kappa_S=1+\lambda(1-r_S)$, respectively. Moreover $r_1=1$, $r_S$ is nonincreasing in $S$, and $r_S\to0$ as $S\to\infty$.
\end{proposition}

\begin{proof}
Let $\hat A_S:=\frac1S\sum_i\varsigma(z_{m_i})-\varsigma(\bar z_S)$ and hold the teacher $t=\sigma(\bar z_S)$ fixed. Since $\nabla\bar z_S=\frac1S\sum_i\nabla z_{m_i}$,
\begin{equation*}
\nabla\,\frac1S\sum_i\mathrm{CE}(t,q_i)\overset{\eqref{eq:ce-softplus}}{=}\frac1S\sum_i\big(\sigma(z_{m_i})-\sigma(\bar z_S)\big)\nabla z_{m_i}=\nabla\hat A_S .
\end{equation*}
Taking $\mathbb E$ over the $S$ masks, which commutes with $\nabla$ because $\Pi$ has finite support, and using $\mathbb E[\bar z_S]=\mu$,
\begin{align*}
\mathbb E\,\nabla\mathcal L_S
&\overset{\eqref{eq:ce-softplus}}{=}\nabla\,\mathbb E\big[\varsigma(\bar z_S)-y\,\bar z_S+\lambda\hat A_S\big]
=\nabla\big[\varsigma(\mu)+A(S)-y\mu+\lambda\big(A-A(S)\big)\big]\\
&\overset{\eqref{eq:ce-softplus}}{=}\nabla\big[\mathrm{CE}(y,\sigma(\mu))+\lambda A+(1-\lambda)A(S)\big],\\
\mathbb E\,\nabla\mathcal L_S'
&\overset{\eqref{eq:ambiguity}}{=}\nabla\big[\mathrm{CE}(y,\sigma(\mu))+A+\lambda\big(A-A(S)\big)\big]
=\nabla\big[\mathrm{CE}(y,\sigma(\mu))+(1+\lambda)A-\lambda A(S)\big].
\end{align*}
For monotonicity, let $\bar z^{(k)}$ be the mean of the $S+1$ logits without the $k$-th; each $\bar z^{(k)}$ has the law of $\bar z_S$, and convexity of $\varsigma$ gives
\begin{equation*}
\varsigma(\bar z_{S+1})=\varsigma\Big(\frac1{S+1}\sum_{k=1}^{S+1}\bar z^{(k)}\Big)\le\frac1{S+1}\sum_{k=1}^{S+1}\varsigma\big(\bar z^{(k)}\big)
\;\Longrightarrow\;A(S+1)\le A(S).
\end{equation*}
Finally $\bar z_1=z_{m_1}$ gives $r_1=1$, and $\bar z_S\to\mu$ almost surely inside a bounded interval fixed by the finite support, so $A(S)\to0$ by dominated convergence.
\end{proof}

Where the second-order expansion \eqref{eq:moments} holds for both $A$ and $A(S)$, $A(S)\approx\tfrac12\varsigma''(\mu)V/S$ and so $r_S\approx1/S$. Two readings follow. At $\lambda=1$ the expected gradient is that of $\mathrm{CE}(y,\sigma(\mu))+A$ for every $S$, the same as ordinary dropout, and $S$ changes only the sampling error of the gradient. At $S=2$, $\lambda=2$ the pooled-logit form has $\kappa_2=2-r_2\approx1.5$ rather than the $\kappa=2$ of its limit; the probability-averaging teacher of \eqref{eq:se-loss} is not covered exactly, since its pool is not $\sigma(\bar z_S)$.

\subsection{Gaussian Surrogate for the Gap}

$A$ is still an expectation over masks. Writing $\varsigma(z_m)=\varsigma(\mu+\delta)$ by Taylor's theorem with integral remainder and averaging over $m$,
\begin{align}
A&=\mathbb E_m\big[\varsigma(\mu+\delta)\big]-\varsigma(\mu)\notag\\
&=\mathbb E_m\Big[\varsigma'(\mu)\,\delta+\delta^{2}\!\int_0^1\!(1-u)\,\varsigma''(\mu+u\delta)\,du\Big]\notag\\
&=\mathbb E_m\Big[\delta^{2}\!\int_0^1\!(1-u)\,\sigma'(\mu+u\delta)\,du\Big],
\label{eq:exact-gap}
\end{align}
where the linear term vanishes because $\mathbb E_m[\delta]=0$, and $\varsigma''=\sigma'$. No approximation has been made: \eqref{eq:exact-gap} reads the gap as the squared perturbation weighted by the Fisher information of the Bernoulli model along the segment from $\mu$ to $z_m$. Since $\sigma'\in(0,\tfrac14]$, that weight is bounded on both sides, and so is $A$.

\begin{theorem}
\label{thm:bounds}
Let $m_\star=\inf\{\sigma'(x):x\in\mathrm{conv}(\mathrm{supp}\,z_m)\}$. Then
\begin{equation}
\tfrac12\,m_\star V\;\le\;A\;\le\;\tfrac18\,V,\qquad V=\mathrm{Var}_m(z_m)=\mathbb E_m[\delta^2].
\label{eq:bounds}
\end{equation}
\end{theorem}

\begin{proof}
\begin{align*}
A&\overset{\eqref{eq:exact-gap}}{=}\mathbb E_m\Big[\delta^{2}\int_0^1(1-u)\,\sigma'(\mu+u\delta)\,du\Big]\\
&\le\tfrac14\,\mathbb E_m\big[\delta^{2}\big]\int_0^1(1-u)\,du=\tfrac18V,
\end{align*}
and replacing $\tfrac14$ by $m_\star$ reverses the inequality.
\end{proof}

The upper bound needs no assumption on the size of $\delta$: it uses only that the gradient of $\varsigma$ is $\tfrac14$-Lipschitz. The lower bound degenerates when the masked logits reach the saturated region of $\sigma$, where $m_\star\to0$; it is informative exactly when they do not.

\begin{figure}[htbp]\centering
\includegraphics{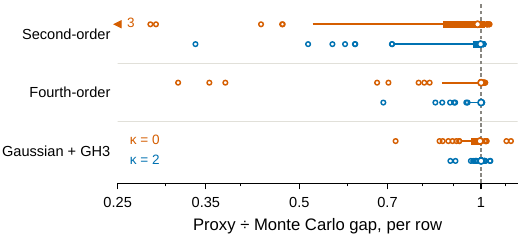}
\caption{Accuracy of three proxies for the gap $A$ of \eqref{eq:exact-gap}, row by row (Criteo validation split). FLMNet towers trained with $\kappa=0$ and $\kappa=2$, three seeds each, on the first $256$ validation rows of each seed, pooled. Each proxy is evaluated at the sample mean and moments of $S=2{,}048$ sampled masks and divided by the Monte Carlo gap $A_{\rm MC}$ from the same masks, so only the treatment of the mask distribution is tested; the closed-form $V$ is tested in Table~\ref{tab:cfvar}, and the full closed-form penalty in Figure~\ref{fig:exp-design}. Second- and fourth-order: \eqref{eq:moments} truncated after the $V$ term and after the $m_4$ term, with sampled $m_3$ and $m_4$. Gaussian + GH3: the penalty term of \eqref{eq:trained}. Thick bars span the 5th to 95th percentiles over rows, thin lines the 1st to 99th, open circles mark the rows beyond, white dots the median, and the dashed line a ratio of $1$; the arrow counts three rows below $0.25$. Training with the second-order truncation in place of GH3 (Criteo, $\kappa=2$, three seeds) changes validation AUC by $0.005\times10^{-4}$ on average, below the seed threshold of $0.94\times10^{-4}$ (Section~\ref{sec:exp-setup}).}\label{fig:surrogate}
\Description{Horizontal interval plot with three rows of proxies (second-order truncation, fourth-order truncation, Gaussian with three-point Gauss--Hermite quadrature) and two sub-rows each for kappa equal to 0 and 2, on a logarithmic axis of the per-row ratio of proxy to Monte Carlo gap from 0.25 to 1.15. All medians lie within 1.5 percent of 1. The second-order truncation has a long left tail, with three rows below 0.25 at kappa 0 and a minimum of 0.34 at kappa 2. The fourth-order truncation has a shorter tail, down to 0.31 at kappa 0 and 0.69 at kappa 2. The Gaussian proxy stays between 0.72 and 1.12 at kappa 0 and between 0.89 and 1.04 at kappa 2.}
\end{figure}

The exact form \eqref{eq:exact-gap} still involves the whole mask distribution. Expanding $\varsigma$ at $\mu$ and using $\mathbb E_m[\delta]=0$,
\begin{equation}
A=\tfrac12\varsigma''(\mu)V+\tfrac16\varsigma'''(\mu)\,m_3+\tfrac1{24}\varsigma''''(\mu)\,m_4+\cdots,\qquad m_k=\mathbb E_m[\delta^k],
\label{eq:moments}
\end{equation}
so $V$ carries the leading term. Truncating after it is the usual analytic treatment of dropout~\cite{wager2013,wei2020}, and it is not the one we adopt: it degrades once $V$ is not small, and truncating at fourth order corrects most rows but not those with large $V$, where it still underestimates $A$ by up to a factor of three (Figure~\ref{fig:surrogate}). Appendix~\ref{app:orders} explains why adding orders need not help.

We instead replace the distribution of $\delta$ by a Gaussian of the same variance and integrate, as in fast dropout~\cite{fastdropout2013}; the difference is that the approximation is applied once, to the logit, whose variance Section~\ref{sec:tower} computes in closed form, rather than propagated layer by layer. Writing $\phi$ for the standard normal density,
\begin{align*}
A&=\mathbb E_m\big[\varsigma(\mu+\delta)\big]-\varsigma(\mu)\\
&\approx\int \varsigma\big(\mu+\sqrt V\,\xi\big)\,\phi(\xi)\,d\xi-\varsigma(\mu)\\
&\approx\sum_{i=1}^{n}\lambda_i\,\varsigma\big(\mu+\xi_i\sqrt V\big)-\varsigma(\mu),
\end{align*}
where the first step replaces the law of $\delta$ by $\mathcal N(0,V)$ and the second applies $n$-point Gauss--Hermite quadrature for the standard normal, with $n=3$. Given $\mu$ and $V$, these are the only two approximations; evaluating $V$ adds a third, the truncation of Section~\ref{sec:closed-form}, which is checked against the exact variance there. Substituting into Theorem~\ref{thm:limit}, the objective we train is
\begin{equation}
\mathcal L\;\approx\;\mathrm{CE}\big(y,\sigma(\mu)\big)+\kappa\Big[\sum_{i=1}^{n}\lambda_i\,\varsigma\big(\mu+\xi_i\sqrt V\big)-\varsigma(\mu)\Big],
\label{eq:trained}
\end{equation}
Because $\sum_i\lambda_i=1$ and $\sum_i\lambda_i\xi_i^2=1$, \eqref{eq:trained} reproduces $\tfrac12\sigma'(\mu)V$ exactly as $V\to0$; unlike that expression, which grows linearly in $V$, it grows no faster than $\sqrt V$, since $\varsigma$ has slope at most one. The bound \eqref{eq:bounds} continues to apply to the exact $A$. Figure~\ref{fig:surrogate} measures how close \eqref{eq:trained} is to \eqref{eq:exact-gap}. On trained towers the Gaussian proxy stays within 28\% of the Monte Carlo gap on every row, while both truncations underestimate it severalfold on rows with large $V$.

At leading order $A$ is proportional to $V$, and for the tower of Section~\ref{sec:tower} $V$ is proportional to $\rho/(1-\rho)$; $\kappa$ and $\rho$ then enter only through the dose $\kappa\rho/(1-\rho)$ and are not separately identifiable. An ablation that removes the distillation term without rescaling $\rho$ reports a change in dose, not in form. Appendix~\ref{app:kappa-rho} gives the computation.

\begin{figure}[htbp]\centering
\includegraphics[width=\linewidth]{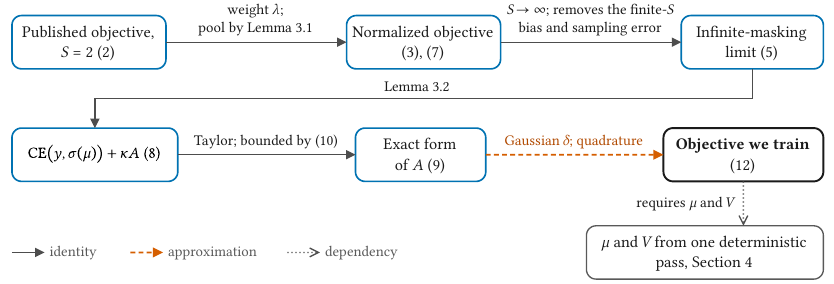}\input{figures/theory_chain_refs}
\caption{From the published objective to the one we train. Solid arrows are identities. The single dashed arrow is the only approximation in this chain: the mask distribution of $\delta$ is replaced by a Gaussian of the same variance, and the resulting integral is evaluated by quadrature. The dotted arrow is not a derivation step: the trained objective needs $\mu$ and $V$, which Section~\ref{sec:tower} computes in one deterministic pass, $\mu$ exactly and $V$ up to a truncation checked in Table~\ref{tab:cfvar}.}
\label{fig:theory-chain}
\Description{Seven boxes: a chain of six in two rows, read left to right, and a seventh below the last, with a legend for three arrow types. The published self-ensemble objective at two masks leads to a normalized objective with an explicit distillation weight, then to its infinite-masking limit; a return line labelled with the ambiguity lemma leads to the label loss plus kappa times the ambiguity gap, then to an exact integral form of that gap. A dashed arrow marks the only approximation, which replaces the mask distribution by a Gaussian and integrates by quadrature, giving the objective that is trained. A dotted arrow, marked as a dependency rather than a derivation step, points from the trained objective to a box stating that the mean logit and its variance come from a single deterministic forward pass.}
\end{figure}

\subsection{The Same Gap One Level Up: Across Seeds}
\label{sec:seedgap}
Section~\ref{sec:limit} penalizes the mask noise of one tower. Towers trained from different random seeds disagree as well, and the same identity accounts for it. Let $\mu_s$ be the mean logit over masks of the tower trained with seed $s$ (for FLMNet, its mask-free forward pass by Proposition~\ref{prop:mean}), let $\bar\mu_K=\frac1K\sum_{s\le K}\mu_s$ be the pooled logit of $K$ seeds, call $\mu_s-\bar\mu_K$ the seed noise of tower $s$, and define
\begin{equation}
A_{\rm seed}(K):=\frac1K\sum_{s=1}^{K}\varsigma(\mu_s)-\varsigma(\bar\mu_K)\;\ge0 .
\label{eq:aseed}
\end{equation}
Lemma~\ref{lem:ambiguity} applies verbatim with the uniform distribution over seeds in place of the mask distribution. Taking $t=y$, which does not depend on $s$, gives
\begin{equation}
\frac1K\sum_{s=1}^{K}\mathrm{CE}\big(y,\sigma(\mu_s)\big)-\mathrm{CE}\big(y,\sigma(\bar\mu_K)\big)=A_{\rm seed}(K).
\label{eq:teacher-gain}
\end{equation}
The logloss that logit pooling gains over the average member is exactly $A_{\rm seed}(K)$: the term $-y\mu$ is linear in the logit and cancels, so the identity holds for either label and on every row. Both levels together give an exact two-level decomposition of the gap between the average over (seed, mask) pairs and the pooled logit,
\begin{equation}
\mathbb E_{s,m}\,\varsigma(z_{s,m})-\varsigma(\bar\mu_K)
=\underbrace{\mathbb E_s\big[\mathbb E_m\varsigma(z_{s,m})-\varsigma(\mu_s)\big]}_{\text{mask gap, the }A\text{ of Section~\ref{sec:limit}, averaged over seeds}}+\underbrace{A_{\rm seed}(K)}_{\text{seed gap}},
\label{eq:two-level}
\end{equation}
which is a telescoping sum. The closed-form penalty \eqref{eq:trained} acts on the first term only, inside each tower. It involves a single tower's own masks and says nothing about $A_{\rm seed}$, so we make no claim about how the penalty changes it.

To move this gain into one tower, a cross-seed student starts from the endpoint of one closed-form tower and regresses the pooled logit,
\begin{equation}
\mathcal L_{\rm KD}=\big(\mu-\bar\mu_K\big)^2,
\label{eq:kd}
\end{equation}
averaged over training rows, where $\mu$ is the student's logit from the same single deterministic pass and $\bar\mu_K$ is computed beforehand and carries no gradient. The student sees neither the label nor the penalty \eqref{eq:trained}. Regressing the logit, rather than matching probabilities, targets the pooled logit $\bar\mu_K$ of \eqref{eq:teacher-gain} directly.

Unlike $A$, the gap $A_{\rm seed}$ has no closed form for our tower. The seed induces a distribution over trained towers through the initialization and, in practice, the data order. Analytic descriptions of the part due to initialization exist only in idealized regimes: in the infinite-width limit, full-batch gradient descent with a small enough step size on squared loss gives predictions that, over random initialization, are distributed as a Gaussian process~\cite{lee2019wide}. Our tower is finite and multiplicatively gated, it is trained on cross-entropy with Adam, minibatches and early stopping, and towers trained from different seeds settle in functionally distinct solutions~\cite{fort2019}, so we do not rely on that limit. Following the common practice of estimating the variance due to optimization by training several seeds on a fixed training set~\cite{neal2018modern}, we estimate $A_{\rm seed}$ from a few trained members, and \eqref{eq:teacher-gain} guarantees that the estimate is the teacher's gain in Logloss (Section~\ref{sec:exp-seed}). Likewise, \eqref{eq:two-level} is an identity between the gaps of fixed predictors. It does not imply that the accuracy gained by closing one gap adds to that gained by closing the other; Section~\ref{sec:exp-seed} measures how far it does.

\section{Single-Tower Architecture for Closed-Form Moment Computation}
\label{sec:tower}
\subsection{Architecture and Gating Dependencies}
Each layer of the tower does three things. A bank of \emph{field-local gate units} reads the embedding: every field's $d$ coordinates feed their own group of ReLU units, and no unit sees more than one field. Before any of this the embedding is recalibrated: each field rescales itself by a linear mix of the others, and the same recalibration is applied again deeper in the chain. A \emph{mixer} maps those units back to the width of the tower, which is where a field's gate acquires influence over the other fields. The result is applied as a \emph{multiplicative update} --- the running representation, which we call the carrier, is rescaled coordinatewise by one plus the gate, so a layer may attenuate or amplify a coordinate while the identity remains available at zero gate. After $L$ such layers a \emph{linear readout} produces the logit. The tower departs from QNN-$\alpha$ in where its gate reads from, in the field-local form of the gate, in the mixer, and in the recalibration; only the first decides whether the closed form is exact, and we return to it below.

Concretely, let $h=Cd$ be the tower width, let $M$ be the number of gate units per tower coordinate, so that one layer holds $hM$ units arranged in $C$ groups of $hM/C$, one group per field, and let the chain start from the recalibrated embedding
\[
x_0=X\odot\big(1+\beta^{\mathrm{fg}}\odot W^{\mathrm{fg}}X\big),\qquad
\big(W^{\mathrm{fg}}X\big)_{f,q}=\sum_{f'}W^{\mathrm{fg}}_{f,f'}X_{f',q},
\]
in which a single $C\times C$ table mixes fields within each embedding dimension and $\beta^{\mathrm{fg}}$ is a per-field scalar gain. This field-mixing form follows the field gate in FinalNet's released implementation~\cite{final2023}; we use the additive residual rather than the concatenating one, and set $\beta^{\mathrm{fg}}=0$ at initialization so that the recalibration starts as the identity. We use $H^{\mathrm{fg}}=2$ tables per application: the $d$ embedding coordinates are split into $H^{\mathrm{fg}}$ contiguous groups and each group carries its own $C\times C$ table, so the groups never mix and the cost in multiply--adds is unchanged. The recalibration is applied at the inputs of the first $D^{\mathrm{fg}}=2$ layers with independent parameters, each application reading the original embedding $X$; applications after the first multiply the carrier only and fold into the preceding $1+g_l$. For $l=1,\dots,L$,
\begin{equation}
\begin{aligned}
r_l&=\mathrm{ReLU}\big(\mathbf G_lx_0+\mathbf b_l\big)\in\mathbb R^{hM},\qquad g_l=U_lr_l+c_l\in\mathbb R^{h},\\
x_l&=x_{l-1}\odot(1+g_l),\qquad z=b_0+\langle w,x_L\rangle.
\end{aligned}
\label{eq:tower}
\end{equation}
Here $\mathbf G_l$ is block diagonal with one block per field --- this is what makes the units field-local --- and $U_l\in\mathbb R^{h\times hM}$ is the mixer. Each block maps one field's $d$ coordinates to that field's $hM/C$ gate units, so a layer holds exactly as many ReLU units as a dense gate of width $hM$ while using $C$ times fewer parameters. Appendix~\ref{app:arch} lists the values of $L$ and $M$ on each benchmark.

\begin{figure}[htbp]
\centering
\includegraphics[width=\linewidth]{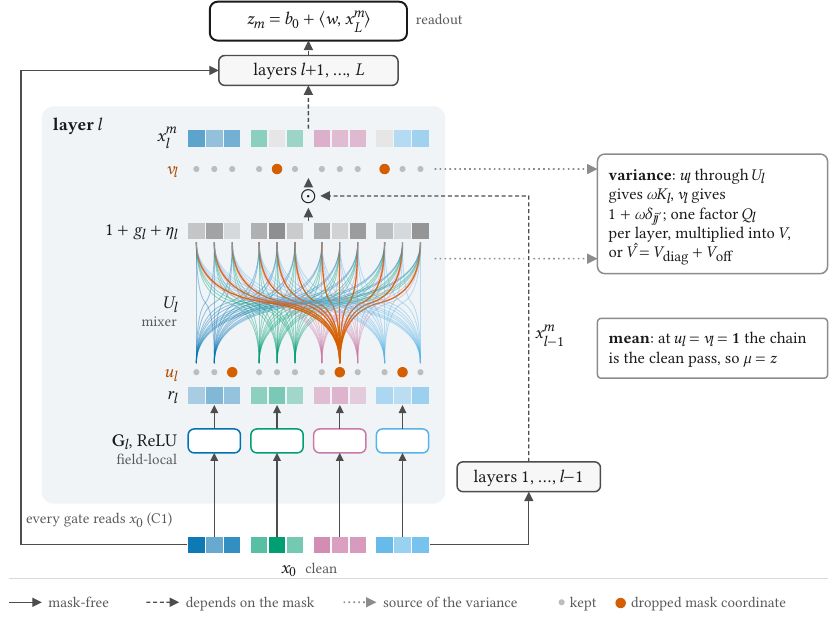}
\caption{Where the masks act in layer $l$ of the tower, drawn for $C=4$ fields with $d=3$ and $M=1$. Each strip is a vector with one cell per coordinate or gate unit, colored by field; the gate output $1+g_l+\eta_l$ is gray, and all shades are illustrative. Solid arrows carry mask-free quantities: the clean embedding $x_0$, which the gate of every layer reads, and the gate units $r_l$, which by (C1) are the same for every mask. Each block of $\mathbf G_l$ reads a single field. The two rows of dots are the mask factors, $u_l$ on the gate units and $v_l$ on the layer output, and the dropped coordinates are orange; dashed arrows carry quantities that depend on the mask. The mixer $U_l$ connects every gate unit to all $h$ tower coordinates, so dropping one unit (orange curves) perturbs $\eta_l$ in every coordinate at once. Its connections are drawn as curves rather than arrows; they carry $u_l\odot r_l$ and therefore depend on the mask. The carrier $x^m_{l-1}$ enters along the right edge, and $\odot$ is its coordinatewise product with the gate output. Gray dotted arrows mark the sources of the variance: $u_l$ through $U_l$ contributes $\omega K_l$ and $v_l$ the factor $1+\omega\,\delta_{jj'}$ in $Q_l$ of \eqref{eq:KQ}. Setting every mask to its mean recovers the clean forward pass, which is the mean logit.}
\Description{A diagram of one tower layer, drawn from bottom to top. The clean embedding x zero, four fields of three coordinates each, feeds four field-local gate blocks G l, one per field, which give the gate units r l. A row of mask dots u l drops some units, and the mixer U l connects every remaining unit to all twelve tower coordinates with curves colored by field; the curves from one dropped unit are highlighted. The result, one plus g l plus eta l, multiplies the carrier x m l minus one, which arrives along the right edge from layers 1 to l minus 1; a second row of mask dots, v l, then drops coordinates, giving x m l, which continues through layers l plus 1 to L to the readout z m. A line on the left edge carries x zero to the gates of the later layers. Dotted arrows from the mixer and from v l lead to a box stating that the variance collects omega K l from u l and one plus omega delta from v l, one factor per layer; another box states that with all masks at one the chain is the clean pass, so the mean equals z.}
\label{fig:mask-flow}
\end{figure}

Dropout adds a second coordinatewise factor to this computation, and since both factors can suppress a coordinate, either could be called a mask. They differ in origin: the gate is computed from the input by learned weights and can attenuate coordinates that carry little information, whereas dropout zeroes coordinates at random, independently of the input. We therefore reserve \emph{mask} for the random factor, the $m$ of Section~2, and \emph{gate} for the learned one. The mask enters at the two points used by the backbone: a factor $u_l$ on the gate units $r_l$ and a factor $v_l$ on the layer output $x_l$ (Figure~\ref{fig:mask-flow}), with every coordinate independent and inverted-scaled as in Section~2, so that $\mathbb E_m[u_l]=\mathbb E_m[v_l]=\mathbf 1$. Neither factor reaches the gate input. Under a mask $m$, layer $l$ therefore computes
\begin{equation}
\label{eq:tower-masked}
x^m_l=v_l\odot x^m_{l-1}\odot\big(1+g_l+\eta_l\big),\qquad
\eta_l=U_l\big((u_l-1)\odot r_l\big),\qquad
z_m=b_0+\langle w,x^m_L\rangle,
\end{equation}
starting from $x^m_0=x_0$. We isolate three properties of this computation that together suffice for the closed form; we do not claim that any of them is necessary.
\begin{itemize}
\item[(C1)] the gate input is unmasked, so $r_l$ is the same for every $m$;
\item[(C2)] the readout is affine in $x^m_L$;
\item[(C3)] each layer rescales the carrier coordinatewise, $x^m_l=x^m_{l-1}\odot\pi_l$, by a factor $\pi_l=v_l\odot(1+g_l+\eta_l)$ that is affine in $u_l$ and in $v_l$.
\end{itemize}

The recalibration is a function of $X$ alone and is therefore constant under the mask; it changes $g_l$ and the coefficients $a_j$, but leaves (C1)--(C3) intact. Unrolling (C3) from $x_0$ and applying (C2) gives, with $a_j=w_jx_{0,j}$,
\begin{equation}
\label{eq:zm-product}
z_m=b_0+\sum_ja_j\prod_{l=1}^{L}\pi_{l,j}.
\end{equation}

\begin{proposition}
\label{prop:mean}
Let the mask coordinates be independent with mean one. Under (C1)--(C3), $\mu=\mathbb E_m[z_m]=z$ for every input $X$ and every parameter $\theta$.
\end{proposition}

\begin{proof}
\begin{align*}
\mu
&\overset{\eqref{eq:zm-product}}{=}b_0+\sum_ja_j\,\mathbb E_m\Big[\prod_{l=1}^{L}\pi_{l,j}\Big]
\overset{(\mathrm a)}{=}b_0+\sum_ja_j\prod_{l=1}^{L}\mathbb E_m[v_{l,j}]\,\big(1+g_{l,j}+\mathbb E_m[\eta_{l,j}]\big)\\
&\overset{(\mathrm b)}{=}b_0+\sum_ja_j\prod_{l=1}^{L}\big(1+g_{l,j}\big)
\overset{\eqref{eq:tower}}{=}b_0+\langle w,x_L\rangle=z.
\end{align*}
Step (a): by (C1), $\pi_{l,j}$ depends only on the masks $(u_l,v_{l,j})$ of layer $l$, so the $L$ factors are independent, and within a factor $v_{l,j}$ is independent of $\eta_{l,j}$. Step (b): $\mathbb E_m[v_{l,j}]=1$, and $\mathbb E_m[\eta_l]=U_l\big((\mathbb E_m[u_l]-1)\odot r_l\big)=0$ because $\eta_l$ is linear in $u_l$ with the $m$-free coefficient $r_l$.
\end{proof}

By \eqref{eq:zm-product}, each mask coordinate enters $z_m$ in at most one factor, and there to the first power. Hence $z_m$ is multilinear in the mask coordinates, and the clean logit $z$ is its value at the mean mask.

None of the three properties is redundant: dropping any one of them, while the other two still hold, can break the identity. A strictly convex, hence non-affine, readout makes $\mu$ exceed $z$ by Jensen's inequality, and a layer that applies a biased ReLU to the masked carrier can do the same. Nor are the three jointly necessary: a gateless network of dense linear layers with dropout on each layer output also has $\mu=z$, because every mask coordinate still enters $z_m$ at most once and to the first power, yet its layers mix coordinates and so violate (C3). (C1) is the property most easily lost in practice. Suppose the gate of layer $l\ge2$ reads the masked carrier $x^m_{l-1}$. Then $v_{l-1}$ scales both the carrier and the gate input of layer $l$, so it generically enters $z_m$ to the second power, even when the gate is linear. Because $\mathbb E_m[v_{l-1,j}^2]=1/(1-\rho)\neq1$ for $0<\rho<1$, the mean then in general differs from the clean forward pass. (C1) asks only that the gate read something unmasked; the input $x_0$ and the clean $x_{l-1}$ both qualify. A recursive backbone such as QNN-$\alpha$ can therefore meet (C1) by computing its gates on a clean forward pass, so that the analogues of Proposition~\ref{prop:mean} and Theorem~\ref{thm:exactvar} hold. The mean and variance they give, however, belong to members whose gates read the clean pass, not to the members QNN-$\alpha$ trains, in which each gate reads that member's masked carrier. It is (C1), not the field-local block structure, that makes $\mu=z$ an identity rather than an approximation.

At $L=1$ these conditions are classical: the identity of Wager et al.~\cite{wager2013} holds exactly under $\mathbb E[\tilde x]=x$, and Wang and Manning~\cite{fastdropout2013} give the first two moments of a single masked linear map in closed form. Ma et al.~\cite{ma2017} name the quantity we set to zero the \emph{inference gap} and give a layerwise sufficient condition --- each layer affine in its masked input --- together with bounds for the deep case; (C3) is weaker in the way that matters, since a layer may be nonlinear in $x$ while remaining linear in the masks of that layer, and independence across layers assembles the whole into a multilinear polynomial. Multilinear extensions are themselves standard; what (C1)--(C3) buy is that the network's own clean forward pass \emph{is} that extension evaluated at the retention probabilities, and that the variance admits the exact recursion of Section~\ref{sec:closed-form}.

\subsection{Closed-Form Mean and Variance}
\label{sec:closed-form}

Proposition~\ref{prop:mean} gives the mean for free; the variance is what the penalty of Section~\ref{sec:limit} actually needs. Because the masks of distinct layers are independent and each layer contributes a single factor to every product, second moments multiply layerwise and the variance closes in one pass over the layers. Write $\omega=\rho/(1-\rho)$ for the variance of a single mask factor, let $F_l$ be the recalibration factor multiplying the output of layer $l$ --- a deterministic function of $X$, equal to one where no recalibration is applied --- and absorb it into the layer as $1+g_l\leftarrow F_l\odot(1+g_l)$ and $U_l\leftarrow\mathrm{diag}(F_l)U_l$, which Appendix~\ref{app:absorb} shows is an exact relabelling rather than an approximation. Collect the one-layer second moments of the gate fan-out in
\begin{equation}
\label{eq:KQ}
K_l=U_l\,\mathrm{diag}\big(r_l^{\odot2}\big)\,U_l^{\top}\in\mathbb R^{h\times h},\qquad
Q_{l,jj'}=\big(1+\omega\,\delta_{jj'}\big)\Big[(1+g_{l,j})(1+g_{l,j'})+\omega K_{l,jj'}\Big].
\end{equation}

\begin{theorem}
\label{thm:exactvar}
Under the conditions of Proposition~\ref{prop:mean}, the mask variance of the logit is
\begin{equation}
\label{eq:exactvar}
V=\sum_{j,j'}a_ja_{j'}\left[\prod_{l=1}^{L}Q_{l,jj'}-\prod_{l=1}^{L}(1+g_{l,j})(1+g_{l,j'})\right].
\end{equation}
\end{theorem}

\begin{proof}
With $z_m=b_0+\sum_ja_j\prod_l\pi_{l,j}$ as in \eqref{eq:zm-product},
\begin{align*}
\mathbb E_m\big[(z_m-b_0)^2\big]
&=\sum_{j,j'}a_ja_{j'}\,\mathbb E_m\Big[\prod_{l=1}^{L}\pi_{l,j}\pi_{l,j'}\Big]
\overset{(\ast)}{=}\sum_{j,j'}a_ja_{j'}\prod_{l=1}^{L}Q_{l,jj'},\\
\big(\mathbb E_m[z_m-b_0]\big)^{2}
&\overset{(\dagger)}{=}\Big(\sum_ja_j\prod_{l=1}^{L}(1+g_{l,j})\Big)^{2}
=\sum_{j,j'}a_ja_{j'}\prod_{l=1}^{L}(1+g_{l,j})(1+g_{l,j'}),
\end{align*}
where $(\ast)$ uses independence across layers together with $\mathbb E_m[\pi_{l,j}\pi_{l,j'}]=Q_{l,jj'}$, and $(\dagger)$ is Proposition~\ref{prop:mean}. Subtracting gives \eqref{eq:exactvar}.
\end{proof}

Equation~\eqref{eq:exactvar} is exact and all-order: it retains every cross term between the two mask families. It is also an $h\times h$ table per layer, two orders of magnitude more expensive than a forward pass, so it cannot be trained with. We use it as the ruler against which the trained form below is checked.

What the exact variance contains is easiest to see through the Hoeffding decomposition, which for a multilinear polynomial is finite and exact,
\begin{equation}
\label{eq:hoeffding}
V=\sum_{T\neq\emptyset}\omega^{|T|}\big(\partial_Tz\big)^2,
\end{equation}
the sum running over sets $T$ of mask coordinates, with $\partial_Tz$ the corresponding mixed derivative of $z_m$, evaluated at the all-ones mask. The term indexed by $T$ is the share of the variance produced by dropping all of $T$ at once that no proper subset of $T$ accounts for. Terms with $|T|=1$ are exactly the delta-method variance used by Wei et al.~\cite{wei2020}. Everything we add lies in $|T|\ge2$, and such terms are nonzero here because a coordinate's contribution is a product across layers: dropping a unit in one layer rescales the effect of dropping a unit in another, so simultaneous drops do not act additively.

\begin{proposition}
\label{prop:cheapvar}
Let $\widehat V$ be obtained from \eqref{eq:exactvar} by keeping the terms with $j=j'$ to all orders in $\omega$ and truncating those with $j\neq j'$ at first order. Then $\widehat V=V_{\mathrm{diag}}+V_{\mathrm{off}}$ with
\begin{align}
V_{\mathrm{diag}}&=\sum_ja_j^2\Big[(1+\omega)^L\textstyle\prod_l\big((1+g_{l,j})^2+\omega K_{l,jj}\big)-P_j^2\Big],\label{eq:vdiag}\\
V_{\mathrm{off}}&=\omega\sum_l\Big[\big\|r_l\odot U_l^{\top}(a\odot P_{-l})\big\|^2-\textstyle\sum_ja_j^2P_{-l,j}^2K_{l,jj}\Big],\label{eq:voff}
\end{align}
where $P_j=\prod_l(1+g_{l,j})$ and $P_{-l,j}=\prod_{l'\neq l}(1+g_{l',j})$, and $V-\widehat V$ consists exactly of the terms of \eqref{eq:exactvar} with $j\neq j'$ of order $\omega^2$ and above.
\end{proposition}

The proof is in Appendix~\ref{app:cheapvar}; the only step worth naming is that the off-diagonal first-order terms collapse into a perfect square, which is what makes $V_{\mathrm{off}}$ computable as a norm rather than as a sum over pairs. Evaluating $\widehat V$ therefore costs two additional matrix multiplications per layer (Appendix~\ref{app:moments}), each of the same shape as the gate's own.

The vector $U_l^{\top}(a\odot P_{-l})$ is the sensitivity of the logit to each gate unit of layer $l$. Dropping one such unit perturbs every coordinate's gate at once, and those perturbations add coherently in the logit; $V_{\mathrm{off}}$ is the variance they contribute. It is precisely the contribution that a mean-field treatment of the same network discards by construction.

Three properties of the numerical form are load-bearing. First, $V_{\mathrm{diag}}$ is accumulated by the recursion
\begin{equation}
\label{eq:vdiag-rec}
D_l=B_lD_{l-1}+\Pi_{l-1}\big(B_l-C_l\big),\qquad
B_{l,j}-C_{l,j}=\omega(1+g_{l,j})^2+(1+\omega)\,\omega K_{l,jj},
\end{equation}
in the notation of Appendix~\ref{app:cheapvar}, whose three terms are nonnegative, so $V_{\mathrm{diag}}\ge0$ holds by construction. Evaluating the same quantity as a difference of products instead subtracts two nearly equal large numbers when $\omega$ is small; an earlier implementation of ours did so, returned negative variances, and cost four training runs before the cause was found. Second, $P_{-l}$ is formed from prefix and suffix products rather than by dividing $P_j$, since $1+g_{l,j}$ may vanish or change sign. Third, $V_{\mathrm{off}}$ is not clamped: it sums covariances between distinct coordinates and is not guaranteed nonnegative, and on a trained model its share of $\widehat V$ has median $5.2\%$ and first percentile $-1.8\%$. We report the examples on which $\widehat V<0$ rather than truncating them; there were none among the $24{,}576$ evaluated here.

Besides $\widehat V$ we check a coarser truncation, $V_{\mathrm{lin}}$, which keeps every term of first order in the gate-unit masks $u_l$, the fan-out $V_{\mathrm{off}}$ included, together with all orders of the output masks $v_l$ alone, and drops every higher-order term that involves some $u_l$. Both truncations can be checked against the exact variance: on small towers by enumerating all $2^{L(hM+h)}$ masks, and on trained models by evaluating \eqref{eq:exactvar} per example. Table~\ref{tab:cfvar} collects the result.

\begin{table}[htbp]\centering\small
\caption{Accuracy of the two truncations against the exact variance \eqref{eq:exactvar}. Left: all masks enumerated, with $W^{\mathrm{fg}},\beta^{\mathrm{fg}}$ randomly perturbed. Right: six trained checkpoints, one per seed, on the same $4096$ validation rows each, in float64; medians over the pooled $24{,}576$ rows, with the $1$st and $99$th percentiles in brackets.}\label{tab:cfvar}
\begin{tabular}{p{.30\linewidth}p{.26\linewidth}p{.32\linewidth}}
\toprule
 & Small towers, exhaustive & Trained checkpoints \\ \midrule
$\widehat V/V$ & $0.99982$--$1.00049$ & $0.99965$ $[0.99808,1.00014]$ \\
First order in $u_l$, $V_{\mathrm{lin}}/V$ & $0.86625$--$0.99910$ & $0.91955$ $[0.85144,0.95966]$ \\
\bottomrule\end{tabular}\end{table}

\FloatBarrier

Truncating at first order in $u_l$ costs $8.0\%$ of the variance and $8.1\%$ of $A$ at the median of a trained model; discarding only the $\omega^2$ remainder costs $0.035\%$ and $3.5\times10^{-4}$ respectively. We train with $\widehat V$ throughout; whether either gap would change accuracy is not tested here.
\subsection{Training and Deterministic Inference}
\label{sec:train-infer}
Training draws no masks. Each optimizer step runs the tower \eqref{eq:tower} once, unmasked, and that single pass supplies everything the objective \eqref{eq:trained} requires: the logit $z$, which equals the mean $\mu$ by Proposition~\ref{prop:mean}, and the gate units $r_l$ and factors $1+g_l$ from which $\widehat V$ of Proposition~\ref{prop:cheapvar} is assembled. The Gaussian surrogate turns $(\mu,\widehat V)$ into the penalty, and the loss is \eqref{eq:trained} averaged over the minibatch. Gradients flow through both $\mu$ and $\widehat V$, so the penalty acts on the gates as well as on the logit. The masking rate $\rho$ enters only through $\omega$. The objective approximates the $S\to\infty$ limit of \eqref{eq:se-loss} with its pool taken in logit space, yet the cost of a step, one pass plus the moments and their gradients, does not grow with $S$, whereas a finite-$S$ self-ensemble runs $S$ masked passes per step. The optimizer, schedule and weight averaging are given in Appendix~\ref{app:train}.

Inference is the same unmasked pass with the penalty switched off. The prediction is $\sigma(z)$ from the delivered weights, and neither $\widehat V$ nor any mask is evaluated. The deployed model is thus an ordinary feed-forward network whose logit is, by Proposition~\ref{prop:mean}, exactly the mean of the masked logits at the same weights, obtained without sampling or averaging at test time.

Cross-seed distillation (Section~\ref{sec:seedgap}) adds a second training stage but leaves inference unchanged. $K$ towers are first trained independently as above, and their pooled logit $\bar\mu_K$ is computed once on the training rows and stored. A student initialized from one of these towers is then trained on \eqref{eq:kd} alone, without the label or the penalty. The student has the same architecture and is deployed as a single tower, so the procedure costs $K$ tower trainings followed by one student run, and still one forward pass per prediction; Section~\ref{sec:exp-summary} reports both costs on Criteo.

\section{Experiments}
\label{sec:exp}

We evaluate FLMNet on five standard benchmarks against seventeen established models and, under a common protocol, report the highest accuracy on every benchmark. Section~\ref{sec:exp-setup} describes the setup and Section~\ref{sec:exp-main} the main comparison; the remaining subsections ask where this accuracy comes from. We start with the tower (Section~\ref{sec:exp-tower}): on QNN-$\alpha$, whose gate violates (C1), the closed form falls short of the Monte Carlo gap and training lowers the estimate far more than the gap it estimates; restoring (C1) costs accuracy, which FLMNet recovers. The remaining subsections trace the accuracy beyond the tower to the suppression of the two kinds of noise defined in Sections~\ref{sec:prelim} and~\ref{sec:seedgap}: mask noise, the disagreement among a tower's predictions under different masks, which the closed-form penalty suppresses at the infinite-masking limit, without sampling masks (Section~\ref{sec:exp-mask}); and seed noise, the disagreement among towers trained from different seeds, which survives that limit. Pooling towers trained with different seeds averages seed noise away, and cross-seed distillation carries the resulting gain into a single tower (Section~\ref{sec:exp-seed}). For each source we ask whether suppressing it helps, why, and whether it helps for the reason the theory gives. Section~\ref{sec:exp-summary} closes with consistency across seeds, cost, and a summary of the answers.

\Needspace{8\baselineskip}
\subsection{Experimental Setup}
\label{sec:exp-setup}

We evaluate on five public benchmarks: Criteo,\footnote{\url{https://www.kaggle.com/c/criteo-display-ad-challenge/data}} Avazu,\footnote{\url{https://www.kaggle.com/c/avazu-ctr-prediction/data}} KKBox,\footnote{\url{https://www.kaggle.com/c/kkbox-music-recommendation-challenge}} Frappe\footnote{\url{https://www.baltrunas.info/context-aware}}~\cite{frappe2015} and ML-1M\footnote{\url{https://files.grouplens.org/datasets/movielens/ml-1m.zip}}. Their statistics are collected in Table~\ref{tab:exp-data}. Each benchmark follows an established preparation, with one adjustment on Avazu described below, and every model is trained and evaluated on the same prepared data. Criteo, Avazu, KKBox and Frappe are taken from their BARS \texttt{x1} releases~\cite{bars2022,openbench2021} and processed under the FuxiCTR\footnote{\url{https://github.com/reczoo/FuxiCTR}} requirements, which fix the split, the vocabulary, the field types and the minimum category count before any model is trained. Criteo keeps the split introduced with AFN~\cite{afn2020}, Frappe is used in the click-through form constructed by NFM~\cite{nfm2017}, and KKBox is split at random in an 8:1:1 ratio. ML-1M has no BARS CTR release, so we follow the preparation of AutoInt~\cite{autoint2019}, which QNN~\cite{qnn2025} also uses: we drop ratings of 3, label ratings above 3 as positive, and split the remaining 739,012 ratings at random in an 8:1:1 ratio. The released validation and test splits of Avazu differ in positive rate (16.8\% versus 15.2\%), which skews model comparison, so we keep its training split and re-split the pooled validation and test data at random into parts of the original sizes. Vocabularies are fitted on the training split alone, and categories unseen there are mapped to a reserved entry. Further details, including where our reproductions depart from published values, the Avazu adjustment, and results not shown in the main tables, are given in Appendix~\ref{app:repro}.\par

\begin{table}[htbp]\centering\small

\caption{Dataset statistics.}\label{tab:exp-data}
\setlength{\tabcolsep}{6pt}
\begin{tabular}{cccccc}
\toprule
\textbf{Dataset} & \textbf{\#Instances} & \textbf{\#Fields} & \textbf{\#Training} & \textbf{\#Validation} & \textbf{\#Test} \\ \midrule
\textbf{Criteo} & 46M & 39 & 33M & 8.3M & 4.6M \\
\textbf{Avazu} & 40M & 22 & 28M & 4.0M & 8.1M \\
\textbf{KKBox} & 7.4M & 13 & 5.9M & 0.74M & 0.74M \\
\textbf{Frappe} & 288,609 & 10 & 202,027 & 57,722 & 28,860 \\
\textbf{ML-1M} & 739,012 & 7 & 591,209 & 73,901 & 73,902 \\
\bottomrule\end{tabular}
\end{table}

We compare against seventeen models taken from the BARS leaderboard and from the baseline sets of CETN~\cite{cetn2025} and QNN~\cite{qnn2025}: three basic references (LR, FM, DNN)~\cite{richardson2007,rendle2010,covington2016}; eleven architectures spanning the main families of explicit and implicit feature interaction (IPNN, DeepFM, xDeepFM, AutoInt+, AFN+, DCNv2, EDCN, MaskNet, FinalMLP, EulerNet, ECN)~\cite{ipnn2016,deepfm2017,xdeepfm2018,autoint2019,afn2020,dcnv22021,edcn2021,masknet2021,finalmlp2023,eulernet2023,fcn2026}; and our three principal competitors, which come from two papers: FinalNet in its 2B configuration~\cite{final2023} and QNN-$\alpha$~\cite{qnn2025} with one head and with two heads. Section~\ref{sec:related} describes the closest of them.\par

We report AUC and Logloss, the two metrics conventionally reported on these benchmarks. AUC is the probability that a randomly drawn clicked impression is scored above a randomly drawn unclicked one. It sees only the induced ranking, which is what a deployed ranker actually serves, and is therefore unaffected by class imbalance. Logloss is the binary cross-entropy of the predicted probability itself, so it rewards calibration rather than order alone, and a model can improve one while losing the other. The literature treats an AUC difference of 0.001 as practically significant in CTR prediction~\cite{openbench2021,dcnv22021}; we do not use this threshold to establish statistical significance. Section~\ref{sec:exp-main} instead reports seed-to-seed confidence intervals for the mean RelaImpr, and Section~\ref{sec:exp-summary} compares the leads seed by seed. To summarize results across benchmarks, we also report the relative improvement over DNN~\cite{yan2014cgl,zhou2018din},
\begin{equation}
\mathrm{RelaImpr}=\Bigl(\frac{\mathrm{AUC}_{\mathrm{model}}-0.5}{\mathrm{AUC}_{\mathrm{DNN}}-0.5}-1\Bigr)\times100\%,
\label{eq:relaimpr}
\end{equation}
where $\mathrm{AUC}_{\mathrm{DNN}}$ is the seed-mean test AUC of DNN on the same benchmark. Subtracting the 0.5 that a random ranker attains puts benchmarks with different AUC levels on a common scale.\par

We report two FLMNet models: FLMNet-CF, the tower trained with the closed-form penalty, and FLMNet-KD, a cross-seed student (Section~\ref{sec:seedgap}) that is initialized from one FLMNet-CF tower and trained to match the pooled logit of $K=3$ FLMNet-CF towers, its own starting tower among them. Our towers are trained with Adam, gradient clipping at 10, single precision, and weight decay on the embeddings only. On every benchmark, our towers are evaluated with an exponential moving average (EMA) of their weights, except distilled students, including FLMNet-KD, which are evaluated with their raw weights. On Criteo, Avazu and KKBox the learning rate follows a cosine decay over a fixed epoch budget, with early stopping on the first two, while Frappe and ML-1M keep the FuxiCTR stopping rule. Batch size, learning rate, weight decay and the averaging horizon are chosen on validation for each benchmark and listed in Appendix~\ref{app:train}. Baselines start from their best publicly recorded configurations; Appendix~\ref{app:baselines} lists every change we made to them, and only one, the batch size of QNN-$\alpha$ on Criteo, was chosen on our validation split. We do not apply cosine decay or weight averaging to them, as these need not suit every architecture. Every model is selected on validation only; the checkpoints to be tested are fixed beforehand, and each is evaluated on the test set once. Test results appear only in the comparison with other models (Section~\ref{sec:exp-main}) and its seed-wise comparison; all other analyses in this section use the validation split. We run six seeds for EDCN, FinalMLP, FinalNet, QNN-$\alpha$, ECN and our towers, and three for the remaining baselines; Tables~\ref{tab:exp-main} and~\ref{tab:exp-main-b} give the exceptions. When we compare two single-seed runs, we call a difference distinguishable from seed noise only if it exceeds $2\sqrt2\,s$, where $s$ is the standard deviation over six FLMNet-CF seeds on the same benchmark. We call $2\sqrt2\,s$ the seed threshold; on Criteo it is $0.94\times10^{-4}$ in AUC and $1.0\times10^{-4}$ in Logloss. Applied to means over several seeds, as in most comparisons below, the threshold is conservative. Parameter counts and FLOPs are reported in Section~\ref{sec:exp-summary}.\par

\paragraph{AI tool usage.} We used LLM-based coding assistants (Claude Code and Codex) to implement the model and training code, to run and monitor the experiments, and to generate the analysis scripts and figures. All designs, derivations, hyper-parameter decisions, and interpretations of the results are our own, and we checked every reported number against the archived checkpoints and logs.\par

\paragraph{Code availability.} Our code for the tower, the closed-form penalty, cross-seed distillation and the evaluation protocol is available at \url{https://github.com/ivanna10240823-rgb/flmnet}.\par

\FloatBarrier

\Needspace{8\baselineskip}
\subsection{Predictive Performance}
\label{sec:exp-main}

\begin{figure}[htbp]\centering
\includegraphics[width=\linewidth]{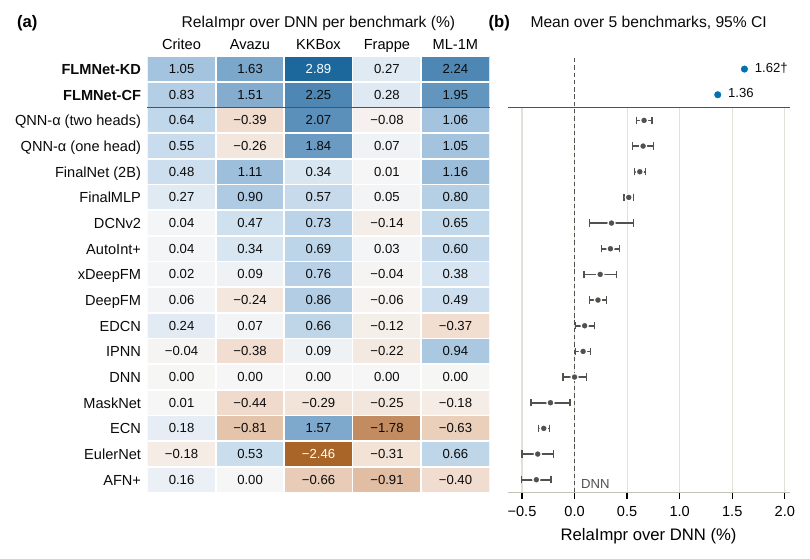}
\caption{Relative improvement over DNN on the five benchmarks, measured by RelaImpr~\eqref{eq:relaimpr}. (a) RelaImpr of each model's seed-mean test AUC on each benchmark; blue marks a gain over DNN, orange a loss. (b) Mean RelaImpr over the five benchmarks with 95\% confidence intervals. The intervals capture seed-to-seed variation only, treating the benchmarks and the DNN reference as fixed, and their degrees of freedom follow the Welch--Satterthwaite approximation; the interval of FLMNet-CF is narrower than its marker. $^\dagger$FLMNet-KD has no interval because its runs are not independent: on every benchmark its students distil from teachers built on overlapping sets of FLMNet-CF towers. LR and FM are omitted here and appear in Tables~\ref{tab:exp-main} and~\ref{tab:exp-main-b}, which also give the seed counts.}
\Description{Panel (a) is a heatmap with one row per model and one column per benchmark, each cell giving the model's relative AUC improvement over DNN in percent, shaded blue when positive and orange when negative. Panel (b) shares the rows and plots each model's mean improvement over the benchmarks as a point with a 95\% confidence interval (FLMNet-KD without one), with a dashed line at zero for DNN. Our two models occupy the top two rows.}
\label{fig:exp-relaimpr}
\end{figure}

On all five benchmarks the two FLMNet models take the top two places in both metrics (Tables~\ref{tab:exp-main} and~\ref{tab:exp-main-b}). Benchmarks differ widely in difficulty, so absolute AUC gaps do not transfer between them, and we compare them through RelaImpr~\eqref{eq:relaimpr}, the AUC gain over DNN divided by DNN's margin above the 0.5 of a random ranker. Among the models published between 2016 and 2026, the best baseline lies only 0.07\% (Frappe) to 2.07\% (KKBox) above DNN, and 0.64\% on Criteo. FLMNet-KD gains 1.4 to 3.8 times as much as that best baseline on every benchmark, for example 1.05\% on Criteo; the factor is largest on Frappe, where the best baseline gains only 0.07\%. Averaged over the five benchmarks (Figure~\ref{fig:exp-relaimpr}), RelaImpr is 1.36\% for FLMNet-CF and 1.62\% for FLMNet-KD, against 0.66\% for QNN-$\alpha$ with two heads, the strongest baseline. Gains of this size are still small in absolute AUC, so we do not judge them from the tables alone; Section~\ref{sec:exp-summary} compares them against a threshold fixed before the test evaluation. Both models run a single tower at inference; the teacher ensemble behind FLMNet-KD is used only in training.\par

\begin{table}[htbp]\centering\footnotesize
\caption{Performance comparison on the two advertising benchmarks, Criteo and Avazu. In each column the three best values are shaded, darker for better: \colorbox{rankhue!32}{\textbf{best}}, \colorbox{rankhue!18}{second}, \colorbox{rankhue!10}{third}. Entries are test means over six seeds for EDCN, FinalMLP, FinalNet, QNN-$\alpha$, ECN and our two towers, and over three seeds for the other baselines; the one exception is FLMNet-KD on Avazu, which has three. All models are trained by us at embedding dimension $d{=}10$, the value BARS tunes for Criteo, and baselines whose original papers used a larger dimension are re-run at $d{=}10$. Criteo follows the BARS \texttt{criteo\_x1} protocol. On Avazu, models are trained on the BARS \texttt{avazu\_x1} training split, and its validation and test splits are pooled and re-split at random into two parts of the original sizes (Appendix~\ref{app:avazu}). On Criteo, our reproductions of FinalNet and FinalMLP fall $3.2\times10^{-4}$ and $4.5\times10^{-4}$ in AUC below the values published on BARS, which reports a single tuned run for each of these two models, whereas every entry here is a seed average; Appendix~\ref{app:gaps} compares our other reproductions with published values.}\label{tab:exp-main}
\setlength{\tabcolsep}{2pt}
\setlength{\extrarowheight}{0pt}
\renewcommand{\arraystretch}{1.0}
\begin{tabular}{>{\centering\arraybackslash}m{20pt}|>{\centering\arraybackslash}m{96pt}|>{\centering\arraybackslash}m{32pt}>{\centering\arraybackslash}m{26pt}|>{\centering\arraybackslash}m{32pt}>{\centering\arraybackslash}m{26pt}}
\hline
\multirow{2}{*}{\textbf{Year}} & \multirow{2}{*}{\textbf{Models}} & \multicolumn{2}{c|}{\textbf{Criteo}} & \multicolumn{2}{c}{\textbf{Avazu}} \\
\cline{3-6}
 & & Logloss$\downarrow$ & AUC$\uparrow$ & Logloss$\downarrow$ & AUC$\uparrow$ \\ \hline
2007 & LR & 0.4609 & 0.7886 & 0.3834 & 0.7463 \\
2010 & FM & 0.4490 & 0.8022 & 0.3788 & 0.7537 \\
2016 & DNN & 0.4382 & 0.8136 & 0.3774 & 0.7578 \\
2016 & IPNN & 0.4384 & 0.8135 & 0.3775 & 0.7569 \\
2017 & DeepFM & 0.4381 & 0.8138 & 0.3775 & 0.7572 \\
2018 & xDeepFM & 0.4382 & 0.8137 & 0.3772 & 0.7581 \\
2019 & AutoInt+ & 0.4381 & 0.8137 & 0.3768 & 0.7587 \\
2020 & AFN+ & 0.4378 & 0.8141 & 0.3771 & 0.7578 \\
2021 & DCNv2 & 0.4383 & 0.8137 & 0.3768 & 0.7591 \\
2021 & EDCN & 0.4376 & 0.8144 & 0.3771 & 0.7580 \\
2021 & MaskNet & 0.4382 & 0.8137 & 0.3781 & 0.7567 \\
2023 & FinalMLP & 0.4375 & 0.8144 & 0.3758 & 0.7601 \\
2023 & EulerNet & 0.4389 & 0.8131 & 0.3766 & 0.7592 \\
2023 & FinalNet (2B) & 0.4368 & 0.8151 & \rankthird{0.3757} & \rankthird{0.7607} \\
2025 & \mbox{QNN-$\alpha$ (one head)} & 0.4372 & 0.8153 & 0.3786 & 0.7572 \\
2025 & \mbox{QNN-$\alpha$ (two heads)} & \rankthird{0.4363} & \rankthird{0.8156} & 0.3777 & 0.7568 \\
2026 & ECN & 0.4378 & 0.8142 & 0.3779 & 0.7557 \\
\hline
ours & FLMNet-CF & \ranksecond{0.4358} & \ranksecond{0.8162} & \ranksecond{0.3748} & \ranksecond{0.7617} \\
ours & FLMNet-KD & \rankfirst{0.4351} & \rankfirst{0.8169} & \rankfirst{0.3745} & \rankfirst{0.7620} \\
\hline
\end{tabular}
\end{table}

\begin{table}[htbp]\centering\footnotesize
\caption{Performance comparison on the three recommendation benchmarks, KKBox, Frappe, and ML-1M. Shading is as in Table~\ref{tab:exp-main}. Entries are test means over six seeds for EDCN, FinalMLP, FinalNet, QNN-$\alpha$, ECN and our two towers, and over three seeds for the other baselines, with these exceptions: FLMNet-CF has eight seeds on ML-1M, and FLMNet-KD has three seeds on every benchmark. All models are trained by us at embedding dimension $d{=}128$ on KKBox ($d{=}16$ for EulerNet), 10 on Frappe and 16 on ML-1M; KKBox and Frappe follow the BARS \texttt{kkbox\_x1} and \texttt{frappe\_x1} protocols. BARS provides no configuration for ML-1M, so the baselines take the architecture hyper-parameters of their BARS \texttt{movielenslatest\_x1} configurations (ECN and EulerNet those of our KKBox runs). On Frappe, FLMNet-CF and FLMNet-KD agree to four decimals in AUC (0.98594 and 0.98590), and the shading follows the unrounded means. QNN-$\alpha$ with four heads, the configuration the QNN authors released for ML-1M, is reported in Appendix~\ref{app:ml1m-qnn4}.}\label{tab:exp-main-b}
\setlength{\tabcolsep}{2pt}
\setlength{\extrarowheight}{0pt}
\renewcommand{\arraystretch}{1.0}
\begin{tabular}{>{\centering\arraybackslash}m{20pt}|>{\centering\arraybackslash}m{96pt}|>{\centering\arraybackslash}m{32pt}>{\centering\arraybackslash}m{26pt}|>{\centering\arraybackslash}m{32pt}>{\centering\arraybackslash}m{26pt}|>{\centering\arraybackslash}m{32pt}>{\centering\arraybackslash}m{26pt}}
\hline
\multirow{2}{*}{\textbf{Year}} & \multirow{2}{*}{\textbf{Models}} & \multicolumn{2}{c|}{\textbf{KKBox}} & \multicolumn{2}{c|}{\textbf{Frappe}} & \multicolumn{2}{c}{\textbf{ML-1M}} \\
\cline{3-8}
 & & Logloss$\downarrow$ & AUC$\uparrow$ & Logloss$\downarrow$ & AUC$\uparrow$ & Logloss$\downarrow$ & AUC$\uparrow$ \\ \hline
2007 & LR & 0.5745 & 0.7678 & 0.3076 & 0.9356 & 0.3509 & 0.8700 \\
2010 & FM & 0.5065 & 0.8299 & 0.1999 & 0.9691 & 0.3261 & 0.8900 \\
2016 & DNN & 0.4813 & 0.8502 & 0.1531 & 0.9846 & 0.3149 & 0.9009 \\
2016 & IPNN & 0.4796 & 0.8506 & 0.1587 & 0.9835 & 0.3113 & 0.9047 \\
2017 & DeepFM & 0.4783 & 0.8532 & 0.1536 & 0.9843 & 0.3120 & 0.9029 \\
2018 & xDeepFM & 0.4784 & 0.8529 & 0.1534 & 0.9844 & 0.3140 & 0.9024 \\
2019 & AutoInt+ & 0.4782 & 0.8527 & 0.1522 & 0.9848 & 0.3150 & 0.9033 \\
2020 & AFN+ & 0.4847 & 0.8479 & 0.2379 & 0.9802 & 0.3204 & 0.8993 \\
2021 & DCNv2 & 0.4789 & 0.8528 & 0.1582 & 0.9839 & 0.3113 & 0.9035 \\
2021 & EDCN & 0.4840 & 0.8526 & 0.1581 & 0.9840 & 0.3170 & 0.8995 \\
2021 & MaskNet & 0.4870 & 0.8492 & 0.1615 & 0.9834 & 0.3196 & 0.9002 \\
2023 & FinalMLP & 0.4801 & 0.8522 & 0.1523 & 0.9848 & 0.3127 & 0.9041 \\
2023 & EulerNet & 0.4935 & 0.8416 & 0.1383 & 0.9831 & 0.3098 & 0.9036 \\
2023 & FinalNet (2B) & 0.4794 & 0.8515 & \rankthird{0.1365} & 0.9846 & 0.3067 & \rankthird{0.9056} \\
2025 & \mbox{QNN-$\alpha$ (one head)} & \rankthird{0.4726} & 0.8567 & 0.1398 & \rankthird{0.9849} & \rankthird{0.3034} & 0.9051 \\
2025 & \mbox{QNN-$\alpha$ (two heads)} & 0.4746 & \rankthird{0.8575} & 0.1426 & 0.9842 & 0.3037 & 0.9052 \\
2026 & ECN & 0.4755 & 0.8558 & 0.2598 & 0.9760 & 0.3177 & 0.8984 \\
\hline
ours & FLMNet-CF & \ranksecond{0.4682} & \ranksecond{0.8581} & \ranksecond{0.1329} & \rankfirst{0.9859} & \ranksecond{0.2981} & \ranksecond{0.9087} \\
ours & FLMNet-KD & \rankfirst{0.4642} & \rankfirst{0.8604} & \rankfirst{0.1324} & \ranksecond{0.9859} & \rankfirst{0.2964} & \rankfirst{0.9099} \\
\hline
\end{tabular}
\end{table}

\FloatBarrier

The two metrics do not always agree on the baselines. On KKBox, EDCN has a higher AUC than DNN (0.8526 against 0.8502) but a worse Logloss (0.4840 against 0.4813), whereas on Frappe FinalNet matches the AUC of DNN (0.9846) at a much lower Logloss (0.1365 against 0.1531). Both FLMNet models improve on DNN in AUC and in Logloss on every benchmark, so their AUC gain does not come at the cost of Logloss.\par

Figure~\ref{fig:exp-relaimpr} shows that no baseline leads on every benchmark and that the baselines do not improve with publication year. The strongest baseline is QNN-$\alpha$ with two heads on Criteo and KKBox, FinalNet on Avazu and ML-1M, and QNN-$\alpha$ with one head on Frappe, where it gains only 0.07\%. ECN (2026) and EulerNet (2023) average $-0.29$\% and $-0.35$\% over the five benchmarks, below DNN. ECN runs a configuration borrowed from Criteo on Avazu and Frappe and from KKBox on ML-1M (Appendix~\ref{app:baselines}), so this average may understate it. FLMNet-CF averages $1.36\pm0.02$\% against $0.66\pm0.07$\% for QNN-$\alpha$ with two heads, and the two intervals do not overlap; they reflect seed variation only, and Section~\ref{sec:exp-summary} checks whether the leads hold across seeds. FLMNet-KD is ahead of FLMNet-CF by 0.12 to 0.64 points of RelaImpr on Criteo, Avazu, KKBox and ML-1M, and level with it on Frappe (0.27\% against 0.28\%). Section~\ref{sec:exp-seed} examines this gap.\par

\FloatBarrier

\Needspace{8\baselineskip}
\subsection{Why a New Tower}
\label{sec:exp-tower}

Section~\ref{sec:tower} shows that the closed form is exact only under (C1), which requires the gate to read an unmasked input. QNN-$\alpha$ violates (C1): its gate reads the masked output of the previous layer and thus inherits every earlier mask, so the layer factors become dependent, and a closed form that treats them as independent misstates both the mean and the variance of QNN-$\alpha$'s logits across masks. On the trained QNN-$\alpha$ towers of Table~\ref{tab:exp-design}, the closed-form mean differs from the sampled mean by more than three standard errors on $76\%$ of the first $256$ validation rows at $\kappa=0$ and on $81\%$ at $\kappa=2$ (computed per seed and averaged over the three seeds), whereas an exact closed form would give about $0.3\%$. The median ratio of the closed-form variance to its Monte Carlo estimate, averaged over seeds in the same way, is $0.58$ at $\kappa=0$ and $0.036$ at $\kappa=2$. Since the penalty is computed from this variance estimate, training can lower the penalty without lowering the true variance.

\begin{figure}[htbp]\centering
\includegraphics[width=\linewidth]{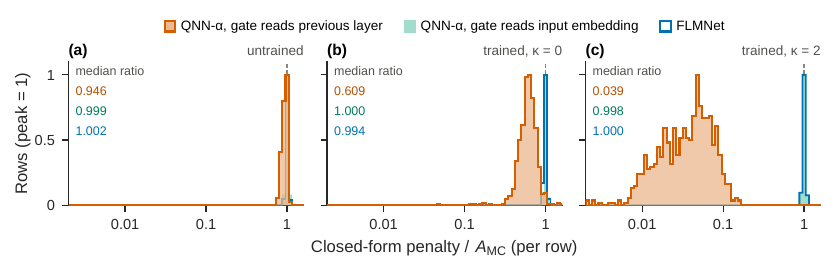}
\caption{Closed-form penalty against the Monte Carlo gap $A_{\rm MC}$, row by row, for the three towers of Table~\ref{tab:exp-fl} on Criteo (validation split). On each of the first $256$ validation rows, the closed-form penalty is divided by $A_{\rm MC}$, the gap $A$ of \eqref{eq:exact-gap} estimated from $S=2{,}048$ sampled masks, as in Table~\ref{tab:exp-design}; each histogram shows the distribution of this ratio on a logarithmic axis, and the dashed line marks a ratio of $1$. (a) Towers at initialization. (b, c) Towers trained with $\kappa=0$ and $\kappa=2$, with the rows of seeds 110000--110002 pooled. Bins are $0.04$ decades wide, and each histogram is scaled to its own peak. Median ratio: median over rows, then mean over seeds. Four rows outside the axis range are counted in the edge bins; at initialization, two rows per tower have a non-positive closed-form penalty and are left out of the histogram but not of the median.}\label{fig:exp-design}
\Description{Three histogram panels, for untrained towers and for towers trained at kappa equal to zero and two, each showing over rows the ratio of the closed-form penalty to the Monte Carlo gap on a logarithmic axis for three towers, with a dashed line at ratio one. For the tower whose gate reads the previous layer, the ratio is already below one at initialization and falls further with training, to a median of 0.039 at kappa equal to two; the other two towers stay near one.}
\end{figure}

Figure~\ref{fig:exp-design} compares, row by row, the closed-form penalty with the Monte Carlo gap $A_{\rm MC}$, the gap $A$ of \eqref{eq:exact-gap} estimated from $S=2{,}048$ sampled masks. When the gate reads the previous layer, the closed-form penalty already falls short of $A_{\rm MC}$ before training (median ratio $0.946$), and training lowers the median ratio to $0.609$ at $\kappa=0$ and $0.039$ at $\kappa=2$. From $\kappa=0$ to $\kappa=2$, $A_{\rm MC}$ drops only from $22.9\times10^{-3}$ to $20.7\times10^{-3}$ (Table~\ref{tab:exp-design}), so the sixteenfold fall in the median ratio comes almost entirely from the closed-form penalty: under the penalty, training lowers the estimate far more than the gap it estimates.

The direct remedy is to restore (C1) by letting the gate of QNN-$\alpha$ read the input embedding. The closed-form mean and variance then become exact, and the penalty lowers $A_{\rm MC}$ from $21.6\times10^{-3}$ to $2.27\times10^{-3}$. The cost is accuracy: reading the input embedding lowers validation AUC from $0.81478$ to $0.81373$ without the penalty ($\kappa=0$), a loss of $10.5\times10^{-4}$, and from $0.81504$ to $0.81453$ with it ($\kappa=2$), a loss of $5.1\times10^{-4}$ (Table~\ref{tab:exp-design}).

\begin{table}[htbp]\centering\footnotesize
\caption{Effect of the gate input on the QNN-$\alpha$ backbone on Criteo (validation split, mean over seeds 110000--110002). Both rows share one backbone ($L=4$, $M=5$, 3.05M tower parameters) and one recipe at $\rho=0.1$, and differ only in what the gate reads: the output of the previous layer, as in QNN-$\alpha$, or the input embedding. Penalty ratio: on each of the first $256$ validation rows, the bracketed penalty of \eqref{eq:trained}, evaluated at the clean logit $z$ and the closed-form variance $\widehat V$, divided by the Monte Carlo gap $A_{\rm MC}$, the gap $A$ of \eqref{eq:exact-gap} estimated from $S=2{,}048$ sampled masks; median over rows, then mean over seeds. A ratio of $1$ means that the closed-form penalty matches $A_{\rm MC}$. $A_{\rm MC}$ columns: averaged over the same rows and then over seeds, in units of $10^{-3}$. Gain: $\kappa=2$ minus $\kappa=0$ under the same seed, in units of $10^{-4}$; the seed threshold of one such difference (Section~\ref{sec:exp-setup}) is $0.94$ for AUC and $1.0$ for Logloss.}\label{tab:exp-design}
\setlength{\tabcolsep}{3pt}
\renewcommand{\arraystretch}{1.0}
\begin{tabular}{c|cc|>{\centering\arraybackslash}p{2.7em}>{\centering\arraybackslash}p{2.7em}|cc|cc}
\hline
\multirow{2}{*}{\textbf{Gate reads}} & \multicolumn{2}{c|}{\textbf{Penalty ratio}} & \multicolumn{2}{c|}{$A_{\rm MC}$} & \multicolumn{2}{c|}{\textbf{AUC}$\uparrow$} & \multicolumn{2}{c}{\textbf{Gain} ($\times10^{-4}$)} \\
\cline{2-9}
 & $\kappa=0$ & $\kappa=2$ & $\kappa=0$ & $\kappa=2$ & $\kappa=0$ & $\kappa=2$ & AUC$\uparrow$ & Logloss$\downarrow$ \\ \hline
previous layer & 0.609 & 0.039 & 22.9 & 20.7 & \textbf{0.81478} & \textbf{0.81504} & $+2.56$ & $-4.23$ \\
input embedding & 1.000 & 0.998 & 21.6 & 2.27 & 0.81373 & 0.81453 & $+7.92$ & $-9.19$ \\
\hline
\end{tabular}
\end{table}

The two gate inputs thus trade accuracy against exactness. Reading the previous layer gives the higher AUC at either value of $\kappa$, whereas reading the input embedding keeps the closed form exact and triples the AUC gain from the penalty, from $2.56\times10^{-4}$ to $7.92\times10^{-4}$. Paired by seed, the gain with the previous-layer gate minus that with the input-embedding gate lies between $-6.2\times10^{-4}$ and $-5.0\times10^{-4}$ in AUC and between $+3.7\times10^{-4}$ and $+6.1\times10^{-4}$ in Logloss; it has the same sign on all three seeds and exceeds the seed threshold several times in magnitude.

Learned gates that attenuate uninformative inputs recur across CTR models and are reported to raise AUC~\cite{masknet2021,gatenet2020,fibinet2019,gdcn2023}, and most of them compute the gate from the input embedding alone. We therefore keep a gate on the input embedding, which preserves (C1), but change its form.

In FLMNet, each gate unit reads a single field of the input embedding, and a learned mixer maps their outputs to the $h$ hidden coordinates of the tower. Section~\ref{sec:tower} specifies the tower in full, including a field recalibration, and derives from it the exact mean and variance on which the penalty rests (Proposition~\ref{prop:mean} and Theorem~\ref{thm:exactvar}). Table~\ref{tab:exp-fl} compares its accuracy, without the penalty, with that of the two QNN-$\alpha$ variants.

\begin{table}[htbp]\centering\footnotesize
\caption{Accuracy without the penalty ($\kappa=0$) on Criteo (validation split, mean over seeds 110000--110002). All towers have $L=4$ layers and $M=5$ branches and are trained with the same recipe at $\rho=0.1$. The first two rows are the rows of Table~\ref{tab:exp-design}; FLMNet differs from the second row in its field-local gates, learned mixer, and field recalibration (Section~\ref{sec:tower}). $\Delta$ is the AUC difference relative to the first row in units of $10^{-4}$; the seed threshold of one difference is $0.94$.}\label{tab:exp-fl}
\setlength{\tabcolsep}{3pt}
\renewcommand{\arraystretch}{1.0}
\begin{tabular}{c|c|c|c|c}
\hline
\textbf{Tower} & \textbf{Gate reads} & \textbf{Tower params} & \textbf{AUC}$\uparrow$ ($\kappa=0$) & $\Delta$ ($\times10^{-4}$) \\ \hline
QNN-$\alpha$ & previous layer & 3.05M & \textbf{0.81478} & -- \\
QNN-$\alpha$ & input embedding & 3.05M & 0.81373 & $-10.5$ \\
FLMNet & input embedding, per field & 3.14M & 0.81469 & $-0.9$ \\
\hline
\end{tabular}
\end{table}

Without the penalty, FLMNet recovers nearly all of the accuracy that QNN-$\alpha$ loses when its gate reads the input embedding (Table~\ref{tab:exp-fl}). Its validation AUC of $0.81469$ is $9.6\times10^{-4}$ above that of this variant and only $0.9\times10^{-4}$ below that of the original QNN-$\alpha$, so on this benchmark the FLMNet tower is about as expressive as one whose gate reads the previous layer. This structure changes the QNN-$\alpha$ variant in three ways at once (field-local gates, a learned mixer, and a field recalibration), and the comparison measures their combined effect without isolating each one. Like the QNN-$\alpha$ variant that reads the input embedding, FLMNet keeps its closed-form mean and variance exact after training, as Theorem~\ref{thm:exactvar} guarantees, and the penalty computed from them matches the Monte Carlo gap closely: its median penalty ratio is $0.994$ at $\kappa=0$ and $1.000$ at $\kappa=2$ (Figure~\ref{fig:exp-design}). Section~\ref{sec:exp-mask} examines the gain from the penalty.

\FloatBarrier

\Needspace{8\baselineskip}
\subsection{Suppressing Mask Noise at the Infinite-Masking Limit}
\label{sec:exp-mask}

Section~\ref{sec:limit} replaces the sampled objective by its infinite-masking limit: the penalty on disagreement among masks is kept, but it is computed in closed form, as the Gaussian surrogate \eqref{eq:trained} of the mask gap built on the exact mean of the masked logit and its closed-form variance $\widehat V$, instead of from a few sampled masks. This leaves open whether removing the sampling error of that estimate costs accuracy, as Wei et al. report for dropout~\cite{wei2020}. We ask whether the closed-form penalty outperforms sampling-based regularizers, whether its lead over dropout survives when dropout is given the same regularization strength (the dose defined below), and when during training the penalty acts. The comparisons with other regularizers, like the component ablation of Section~\ref{sec:exp-seed}, cover the four benchmarks other than ML-1M; the timing experiment uses Criteo alone.

\begin{figure}[htbp]\centering
\includegraphics[width=\linewidth]{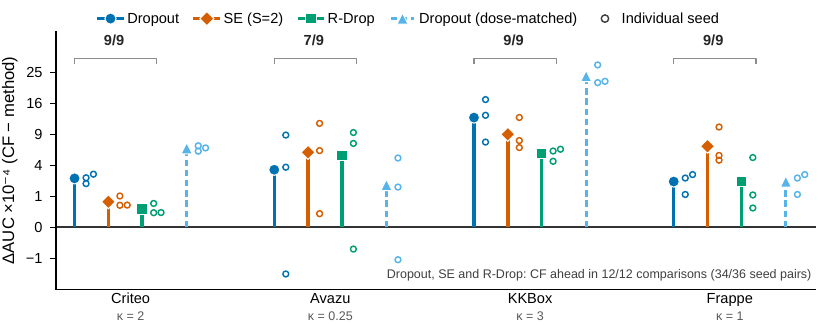}
\caption{Closed-form penalty against sampling-based regularizers on the same tower and backbone (validation split). Each stem shows the AUC of FLMNet-CF at the $\kappa$ it uses on that benchmark (given below it) minus that of a sampling-based regularizer, averaged over three seeds, with runs paired by seed; open circles show the individual seed pairs. From left to right in each group, the regularizers are dropout at $\rho=0.1$, the self-ensemble loss ($S=2$), R-Drop with its weight chosen on validation, and dropout at the dose-matched rate $\rho'$ (dashed). The rate $\rho'$ gives dropout the dose $\kappa\rho/(1-\rho)$ of FLMNet-CF; on Frappe it equals $0.1$, so the dashed stem there reuses the runs of the first stem. The bracketed count above each benchmark is the number of seed pairs, out of the nine under the first three stems, in which the closed form leads. The vertical axis uses a signed square-root scale in units of $10^{-4}$.}\label{fig:exp-dose}

\Description{Single-panel chart with four benchmark groups, Criteo, Avazu, KKBox and Frappe. Each group holds three solid vertical stems for dropout, the self-ensemble loss and R-Drop, followed by a dashed stem for dose-matched dropout. Each stem rises from zero to the mean validation AUC difference between the closed-form penalty and that method, on a signed square-root axis in units of ten to the minus four, with open circles for the three individual seeds. All stems lie above zero; three Avazu seed points lie below zero. A bracket over the three solid stems of each group carries the count of positive seed pairs: 9 of 9, 7 of 9, 9 of 9 and 9 of 9.}\end{figure}

Figure~\ref{fig:exp-dose} compares FLMNet-CF at the $\kappa$ it uses on each benchmark with three sampling-based regularizers on the same tower and backbone: dropout at $\rho=0.1$, whose expected objective is the $\kappa=1$ case of Theorem~\ref{thm:limit}; the self-ensemble loss at $S=2$~\cite{qnn2025}, whose dose differs from that of its limit (Proposition~\ref{prop:finite}); and R-Drop, which penalizes the divergence between two dropout passes without a teacher~\cite{rdrop2021}. The closed form has the higher mean validation AUC in all twelve benchmark--method pairs and leads in 34 of the 36 seed-paired comparisons. Counting each benchmark and seed as one unit, it leads all three methods in 11 of 12 units (one-sided sign test, $p\approx0.003$). Dropout at $\kappa=1$ applies half the dose of FLMNet-CF on Criteo and a third of it on KKBox, so part of the lead over dropout on these two benchmarks may come from the dose rather than from the closed form; the dose-matched comparison below separates the two.

Matching the dose relies on Section~\ref{sec:limit}: to leading order, $\kappa$ and $\rho$ enter the objective only through the dose $\kappa\rho/(1-\rho)$, so dropout matches the dose of FLMNet-CF at $\rho'=2/11$ on Criteo and $\rho'=1/4$ on KKBox (the fourth stem in Figure~\ref{fig:exp-dose}). At this dose the closed form still leads on both benchmarks in every seed pair, by $6.5\times10^{-4}$ on Criteo and $23.9\times10^{-4}$ on KKBox on average, so the dose does not account for its lead. On Avazu ($\rho'=1/37$), the mean lead at the matched dose is $1.9\times10^{-4}$ and the paired differences change sign across seeds, so the dose may account for the lead there.

\begin{figure}[htbp]\centering
\includegraphics[width=\linewidth]{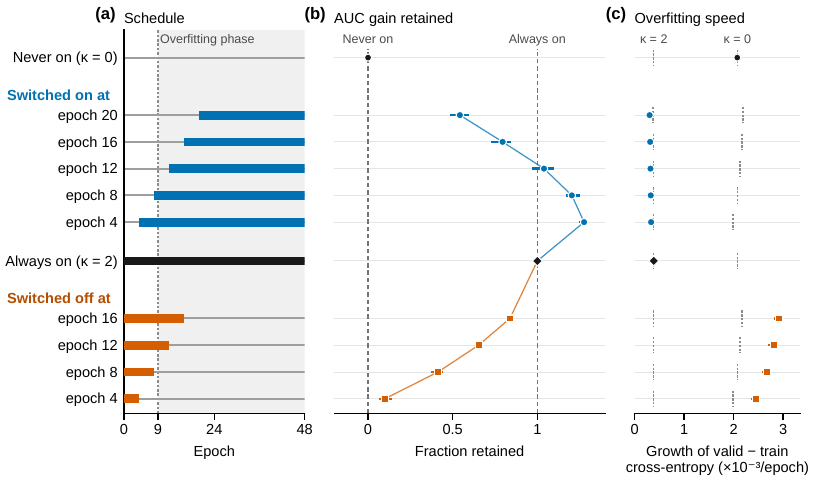}
\caption{Switching the penalty on or off during training (Criteo, constant learning rate, $\kappa=2$, 48 epochs, three seeds; validation split). Each row is one penalty schedule; in (b) and (c), markers give the mean over three seeds and whiskers their range. (a)~Schedule: thick bars mark the epochs with the penalty on; shading marks the epochs after the never-penalized run passes its validation peak (epoch 9). (b)~AUC gain retained: the gain of each run in peak validation AUC over the never-penalized run, divided by that of the always-penalized run, computed per seed. Lines join the means. The always-penalized run is their shared end point, read as onset at epoch 0 on the switch-on line and as no switch-off on the switch-off line; the switch-on line peaks at onset epoch 4. (c)~Growth rate of validation minus training cross-entropy (penalty excluded): the least-squares slope from four epochs after the switch to epoch 48, or from epoch 12 for the never- and always-penalized runs. Grey dotted ticks give the slopes of these two runs ($\kappa=0$ and $\kappa=2$) fitted over the same epochs as each row.}\label{fig:exp-onset}

\Description{Three aligned panels on Criteo with one row per training run: never penalized, penalty switched on at epoch 20, 16, 12, 8 or 4, always penalized, and penalty switched off at epoch 16, 12, 8 or 4. Panel a draws each run's penalty schedule as a bar over the epochs, with the epochs after the never-penalized run's validation peak shaded. Panel b plots, for each run, the fraction of the always-penalized AUC gain that the run retains, as a mean marker with a whisker for the seed range and reference lines at zero and one. A blue line joins the switch-on runs to the always-penalized run and bulges to the right, peaking at about 1.28 for onset at epoch 4; an orange line falls from the always-penalized run through the switch-off runs to about 0.10 at epoch 4. Panel c plots each run's growth rate of validation minus training cross-entropy, with grey reference ticks for the never- and always-penalized runs over the same epochs; switch-on runs sit near 0.3 and switch-off runs between 2.5 and 2.9 thousandths per epoch.}\end{figure}

To see why the closed-form penalty raises validation AUC, we vary when during training it is active. The timing indicates what the penalty does: it holds back overfitting, and in the first epochs it slows learning (Figure~\ref{fig:exp-onset}). On Criteo, we train FLMNet-CF at a constant learning rate, so that the switching epoch is not confounded with the schedule, and switch the penalty ($\kappa=2$) on or off at epoch $t$ (three seeds). Without the penalty, validation AUC peaks around epoch 9 and then declines; the always-penalized run peaks $9.4\times10^{-4}$ higher. Across switch-on runs, the peak validation AUC traces an inverted U in the onset epoch, in every seed: it is highest for onset at epoch 4, $2.6\times10^{-4}$ above the always-penalized run, while onset at epoch 20 keeps about half of the gain of the always-penalized run over the never-penalized one. The earlier the penalty is switched off, the lower the peak, and a switch-off at epoch 8 or earlier keeps less than half of the gain.

Panels~b and~c show that the penalty acts largely on the speed of overfitting. In panel~c, while the penalty is on, the gap between validation and training cross-entropy grows by about $0.3\times10^{-3}$ per epoch whatever the onset epoch, six to seven times slower than in the never-penalized run over the same epochs; once the penalty is switched off, the gap grows by $2.5$ to $2.9\times10^{-3}$ per epoch, faster than in the never-penalized run. The speed of overfitting thus follows whether the penalty is on, not how far training has progressed, and panel~b agrees: the earlier the penalty is switched off, the less of its gain survives. Panel~b also shows a cost early in training. Starting the penalty at epoch 4 retains more gain than applying it from the start, and the always-penalized run trails the never-penalized one from epoch 2 to 10 in every seed, suggesting that in the first epochs, while there is still little overfitting to hold back, the penalty slows the learning of the dominant regularities, as it grows with every contribution to the masked variance (Theorem~\ref{thm:exactvar}). These runs use one benchmark and a constant learning rate rather than the schedule of our main results, and with onsets tested every four epochs they cannot locate the best onset more precisely. Overall, the penalty computed in closed form outperforms the sampling-based regularizers tested here, and it acts mainly by slowing overfitting, in line with the intuition of Section~\ref{sec:limit-inf}. It reduces disagreement among the masks within one tower, but towers trained with different seeds still disagree with one another, and the infinite-masking limit does not remove this disagreement. Section~\ref{sec:exp-seed} suppresses it by distilling across seeds.

\FloatBarrier

\Needspace{8\baselineskip}
\subsection{Suppressing Seed Noise by Cross-Seed Distillation}
\label{sec:exp-seed}

By \eqref{eq:teacher-gain}, a teacher that pools the logits of towers trained with different seeds gains exactly $A_{\rm seed}$ in Logloss over the average member. We show empirically that distillation carries this gain into a single tower, that further training on labels does not reach it, and that it is largely separate from the gain of the penalty, which acts on the disagreement among masks within one tower.

To bring this gain into one tower, we train a student as FLMNet-KD is trained (Section~\ref{sec:seedgap}), from the endpoint of one FLMNet-CF seed, $\theta_0$, toward a teacher that pools the logits of five other seeds. The teacher's logits are computed beforehand for training, and inference uses the student tower alone. On Criteo, this multi-teacher student raises validation AUC by $0.80\times10^{-3}$ over the average member, more than half of the $1.44\times10^{-3}$ that the teacher gains with five towers, and Logloss follows the same pattern (Table~\ref{tab:exp-transfer}). By \eqref{eq:teacher-gain}, the teacher's Logloss gain is the cost of the seed noise that pooling averages away; a student trained toward the pooled logit is therefore drawn away from its own seed noise and toward what the seeds share. Further training on labels does not reach this gain. We continue $\theta_0$ on labels with the same optimizer and learning rate as the students and, like them, without the penalty; validation AUC falls from the first epoch on, since the labels say nothing about how other seeds would predict.

To trace the source of the gain, we run an ablation with a single teacher ($K=1$), in which five students start from $\theta_0$ and each is distilled toward one other seed. Their AUC barely moves: it lies $0.10\times10^{-3}$ above the member mean, close to the seed threshold of $0.94\times10^{-4}$, because a single seed carries idiosyncrasies of its own. These students are useful instead because they let us compare, from the common origin $\theta_0$, the directions toward different seeds. Parameters of different seeds cannot be compared directly, since their initializations differ and their hidden units are not aligned. Each single-teacher student, by contrast, links $\theta_0$ through one continuous training path to a point whose predictions have moved part of the way toward those of another seed, and both ends of the path share the parameterization of $\theta_0$. We next examine the loss along the lines from $\theta_0$ to these endpoints and to that of the multi-teacher student.

\begin{table}[htbp]\centering\small

\caption{Distilling cross-seed teachers into one tower on Criteo (validation split), with rows ordered by gain. The member row is the mean of six FLMNet-CF seeds with EMA weights. Every student starts from one of these seeds, $\theta_0$: the single-teacher row ($K=1$) averages five students, each distilled toward one of the other five seeds, and the multi-teacher student is distilled toward the teacher ensemble, which pools the logits of those five seeds. Students are evaluated on raw weights. The teacher ensemble runs five towers at inference and serves only as a reference. Gains are AUC / Logloss differences from the member mean over all six seeds, in units of $10^{-3}$; the teacher's gain over the mean of its own five members is the $A_{\rm seed}$ of \eqref{eq:teacher-gain}. The starting point $\theta_0$ has a validation AUC $0.03\times10^{-3}$ below the member mean, so the students' AUC gains over their own starting point are larger by this amount.}\label{tab:exp-transfer}

\begin{tabular}{p{.30\linewidth}p{.09\linewidth}p{.11\linewidth}p{.27\linewidth}p{.11\linewidth}}

\toprule
Model & AUC$\uparrow$ & Logloss$\downarrow$ & Gain over direct training & Inference models \\ \midrule

Directly trained member & 0.81585 & 0.43630 & -- & 1 \\

Single-teacher student ($K=1$) & 0.81595 & 0.43616 & $+0.10$ / $-0.14$ & 1 \\

Multi-teacher student ($K=5$) & 0.81665 & 0.43547 & $+0.80$ / $-0.83$ & 1 \\

Teacher ensemble & 0.81729 & 0.43484 & $+1.44$ / $-1.46$ & 5 \\

\bottomrule\end{tabular}\end{table}

\begin{figure}[htbp]\centering

\includegraphics[width=\linewidth]{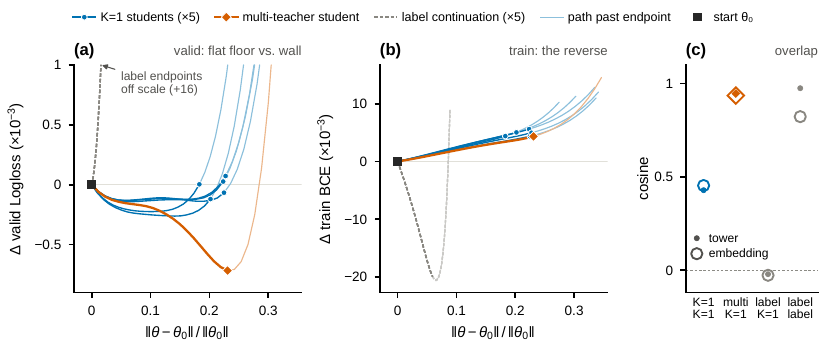}

\caption{The loss surface around one seed endpoint along distillation and label directions (Criteo; fixed subsets of 200{,}000 valid and 200{,}000 train rows). Every run starts from one CF seed endpoint $\theta_0$; the five other seeds are the teachers. In (a) and (b) the horizontal axis is the Euclidean distance from $\theta_0$ over all parameters divided by $\|\theta_0\|$ (dominated by the tower); each curve follows the straight path from $\theta_0$ through one trained endpoint (marker) to 1.5 times its length, drawn lighter beyond the endpoint. (a)~Change in valid Logloss. Along the distillation directions the loss stays level with or below that of $\theta_0$ out to about $0.2\|\theta_0\|$: the multi-teacher student ends at the lowest point of its path ($-0.72\times10^{-3}$), and the five single-teacher ($K{=}1$) students end near the rim of this flat stretch ($-0.12$ to $+0.07\times10^{-3}$). Label-only continuation under five data orders ends only $0.054\|\theta_0\|$ from $\theta_0$, yet $16\times10^{-3}$ higher. (b)~Change in train label cross-entropy along the student directions of (a) and along label continuation: it rises toward the students and falls toward label continuation. (c)~Cosine between parameter displacements from $\theta_0$ after their component along $\theta_0$ is projected out, separately for the tower and the embedding parameters (small filled marker: tower; large open marker: embeddings, drawn in the same column), one column per kind of pair: $K{=}1$ students with each other, the multi-teacher student with the mean $K{=}1$ displacement, label continuation with $K{=}1$ students, and label continuations with each other. Markers give the mean over pairs, 0.43/0.45, 0.95/0.94, $-0.02$/$-0.03$ and 0.98/0.82 (tower/embeddings); the range across pairs is at most 0.04; the dashed line at 0 marks random directions, whose standard deviation is below $10^{-3}$. Parameters are compared only among points that share $\theta_0$ as origin; the other seeds lie about $1.41\|\theta_0\|$ from $\theta_0$ without permutation alignment, an upper bound on their distance after aligning hidden units, and are not drawn. Seed endpoints use EMA weights; students and label continuation use raw weights. In (a) and (c), label continuation is taken at the checkpoint closest to the mean step at which the $K{=}1$ students reach their peak validation AUC; in (b), its line runs through its final checkpoint instead.}\label{fig:exp-seed-basin}

\Description{Three-panel figure, in one row, on the loss surface around one trained seed. Panels a and b share a horizontal axis, the parameter distance from the starting seed relative to its norm, and trace straight paths from the starting seed through each trained endpoint. Panel a shows validation log loss: along the five single-teacher student paths and the multi-teacher student path it stays level or dips slightly over a flat stretch out to about 0.2, with the multi-teacher student ending at its lowest point, about 0.7 thousandths below the start; along the label-only continuation paths it climbs off the scale within a distance of about 0.05. Panel b shows training cross-entropy along the same student paths and along label continuation through its final checkpoint: it rises slowly toward the students and drops steeply toward label continuation. Panel c is a narrow dot plot, one column per kind of pair with tower and embedding values in the same column, of cosines between parameter displacements with the starting-point axis removed, for tower and embedding parameters: about 0.43 between single-teacher students, about 0.95 between the multi-teacher student and the mean single-teacher displacement, about zero between label continuation and the students, and about 0.98 among label continuations.}\end{figure}

\begin{figure}[htbp]\centering

\includegraphics[width=\linewidth]{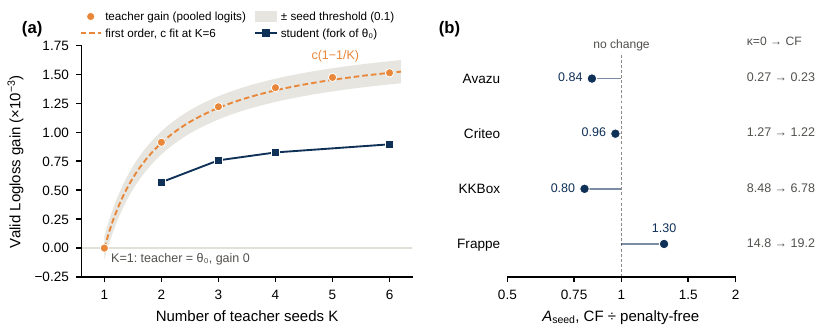}

\caption{Teacher gain against the number of pooled seeds, and seed disagreement with and without the penalty (validation split). (a)~Criteo. Each teacher pools the logits of the first $K$ FLMNet-CF seeds in a fixed order (110000, 110001, \ldots), and its gain, the mean member Logloss minus the teacher Logloss, equals $A_{\rm seed}(K)$. The order starts at $\theta_0$ (seed 110000), so the $K{=}1$ teacher is $\theta_0$ itself, with gain 0; the teacher ensemble of Table~\ref{tab:exp-transfer} instead excludes $\theta_0$. The gain rises from $0.91\times10^{-3}$ at $K=2$ to $1.51\times10^{-3}$ at $K=6$. The dashed curve is $c(1-1/K)$ with $c$ fitted at $K=6$ alone, and the band marks $\pm10^{-4}$, the seed threshold of one Logloss difference; the agreement is consistent with the seeds behaving as interchangeable draws. Students continue from $\theta_0$ toward each teacher; their gain is measured against the EMA weights of $\theta_0$ rather than the member mean and is $0.59$ to $0.62$ times the teacher gain at every $K$ from 2 to 6. Students distilled toward a single other seed appear in Table~\ref{tab:exp-transfer}. (b)~Seed disagreement $A_{\rm seed}$ among three members per benchmark (seeds 110000--110002; Frappe 110008--110010), for FLMNet-CF members divided by that for penalty-free members ($\kappa=0$) with the same seeds. $A_{\rm seed}$ equals the Logloss gain of the teacher that pools these three members; the column on the right gives the two values, penalty-free $\to$ FLMNet-CF, in $10^{-3}$ nats per row. The log axis spans a factor of two either way, and the dashed line at 1 marks no change; each ratio rests on three seeds per side and has no confidence interval. On the same members, the penalty lowers the Gaussian surrogate of the mask gap, which the panel does not show, to $4$--$20\%$ of its penalty-free value.}\label{fig:exp-seed}

\Description{Two-panel figure. Panel a plots the Criteo validation Logloss gain of logit-pooled teachers and of distilled students against the number of teacher seeds K from 1 to 6, with a first-order curve c(1-1/K) fitted at K equals 6 and a band of plus or minus the seed threshold; the K equals 1 teacher is the starting point itself and has zero gain, the teacher points lie on the curve, and student gains rise from about 0.57 to 0.90 times ten to the minus three. Panel b is a lollipop chart, on a log axis from 0.5 to 2, of the closed-form to penalty-free ratio of the seed disagreement among three members, one row per benchmark, with a dashed reference line at 1 and the two values listed on the right in units of ten to the minus three: about 0.27 and 0.23 on Avazu, 1.27 and 1.22 on Criteo, 8.5 and 6.8 on KKBox, and 14.8 and 19.2 on Frappe, so the closed-form to penalty-free ratios are 0.84, 0.96, 0.80 and 1.30.}\end{figure}

The result of this ablation is somewhat unexpected (Figure~\ref{fig:exp-seed-basin}). Along the lines from $\theta_0$ toward the students, validation Logloss stays level or decreases, and the multi-teacher student ends at the lowest point of its line, $0.72\times10^{-3}$ below $\theta_0$. The cross-entropy on training labels (penalty excluded), by contrast, increases along every one of these lines and decreases slightly when they are extended to the opposite side of $\theta_0$ (not drawn). At $\theta_0$, the training labels therefore no longer point toward parameters that generalize better. Label continuation does the opposite: it lowers the training loss and raises validation Logloss by $16\times10^{-3}$ within about $0.05\|\theta_0\|$, less than a third of the distance the students travel. This drift does not depend on the data order. Once their component along $\theta_0$ is projected out, the displacements of label continuation under five data orders point in nearly the same direction, which is nearly orthogonal to the displacements of the students (panel~c). The single-teacher students, although taught by different seeds, share a large common component of their displacements, and the multi-teacher student follows their mean, so pooling the teachers retains this component. Distillation and label training thus move $\theta_0$ along directions that barely overlap. We take this to mean that $\theta_0$ has converged as far as the labels can take it. What the training labels can still supply is specific to individual training examples, the kind of fitting that the penalty holds back within one tower (Section~\ref{sec:exp-mask}), so further label training leads to overfitting. The labels carry no signal toward what the seeds share, which has to be brought in from other seeds. All of these paths start from a single seed on one benchmark.

Finally, the penalty and distillation both counter overfitting, yet their gains largely add up, consistent with each reducing a different disagreement. Distillation draws on the disagreement between seeds. Pooling $K$ seeds gains $A_{\rm seed}(K)$, which grows as $1-1/K$ (Figure~\ref{fig:exp-seed}a), the rate at which averaging cancels independent seed noise: by Theorem~\ref{thm:bounds}, $A_{\rm seed}(K)$ follows the spread of the $K$ member logits, whose expectation for interchangeable members scales as $1-1/K$. The students keep about $0.6$ times this gain at every $K$. The penalty instead reduces the disagreement among masks within one tower: it lowers the Gaussian surrogate of the mask gap \eqref{eq:trained} to $4$--$20\%$ of its penalty-free value on the four benchmarks, while towers trained with it disagree with one another about as much as towers trained without it (Figure~\ref{fig:exp-seed}b). Consistent with this, the two gains combine (Table~\ref{tab:exp-components}). Applied together, they give the best result on nearly every benchmark and metric; on Criteo and KKBox, where each gain alone exceeds the seed threshold, the combination keeps most of their sum, and on Avazu and Frappe the AUC differences among the four rows lie below the seed threshold. The penalty makes the masks within a tower agree, and distillation brings in what the seeds agree on, which the labels cannot reach.

\begin{table}[htbp]\centering\footnotesize
\caption{Contribution of each component to validation AUC and Logloss (mean over three seeds; Frappe uses seeds 110008--110010, the other benchmarks 110000--110002). The penalty column marks whether a model is trained with the closed-form penalty at the $\kappa$ that FLMNet-CF uses on its benchmark. In the distillation rows a single student continues from a member's endpoint toward a logit-pooled teacher of three members, and the penalty column then refers to how those members were trained. On Criteo the last row reuses an earlier run in which each student's teacher pools its own closed-form member with two others drawn at random from seeds 110000--110005 (for seed 110000, seeds 110001 and 110003), whereas in the third row every student's teacher pools the penalty-free members with seeds 110000--110002. Students are evaluated on raw weights, the other rows on EMA weights. On Criteo and KKBox, the last row retains $87\%$ and $80\%$ of the sum of the two single-component AUC gains. Best value per column in bold.}\label{tab:exp-components}
\setlength{\tabcolsep}{2pt}
\setlength{\extrarowheight}{0pt}
\renewcommand{\arraystretch}{1.0}
\begin{tabular}{>{\centering\arraybackslash}m{46pt}|>{\centering\arraybackslash}m{52pt}|>{\centering\arraybackslash}m{28pt}>{\centering\arraybackslash}m{30pt}|>{\centering\arraybackslash}m{28pt}>{\centering\arraybackslash}m{30pt}|>{\centering\arraybackslash}m{28pt}>{\centering\arraybackslash}m{30pt}|>{\centering\arraybackslash}m{28pt}>{\centering\arraybackslash}m{30pt}}
\hline
\multirow{2}{=}{\textbf{Closed-form penalty}} & \multirow{2}{=}{\textbf{Cross-seed distillation}} & \multicolumn{2}{c|}{\textbf{Criteo}} & \multicolumn{2}{c|}{\textbf{Avazu}} & \multicolumn{2}{c|}{\textbf{KKBox}} & \multicolumn{2}{c}{\textbf{Frappe}} \\
\cline{3-10}
 & & AUC$\uparrow$ & Logloss$\downarrow$ & AUC$\uparrow$ & Logloss$\downarrow$ & AUC$\uparrow$ & Logloss$\downarrow$ & AUC$\uparrow$ & Logloss$\downarrow$ \\ \hline
-- & -- & 0.81469 & 0.43732 & 0.76194 & 0.37471 & 0.85453 & 0.47294 & 0.98670 & 0.13115 \\
$\surd$ & -- & 0.81586 & 0.43631 & 0.76182 & 0.37479 & 0.85821 & 0.46808 & 0.98703 & 0.13087 \\
-- & $\surd$ & 0.81569 & 0.43633 & 0.76212 & \textbf{0.37443} & 0.85824 & 0.46735 & 0.98671 & 0.13066 \\
$\surd$ & $\surd$ & \textbf{0.81658} & \textbf{0.43555} & \textbf{0.76213} & 0.37451 & \textbf{0.86046} & \textbf{0.46417} & \textbf{0.98709} & \textbf{0.12709} \\
\hline
\end{tabular}
\end{table}

\Needspace{8\baselineskip}
\subsection{Consistency Across Seeds, Cost, and Summary}
\label{sec:exp-summary}

The leads in Section~\ref{sec:exp-main} are consistent across seeds. Before the test evaluation, we fixed for each benchmark the opponent, the strongest baseline on validation, and a threshold on the mean seed-wise difference in test AUC: twice its standard error, or a preset floor if larger (Appendix~\ref{app:protocol}). FLMNet-CF exceeds this threshold on every benchmark, by a factor of $1.9$ to $8.7$, and leads in every seed pair wherever seeds are paired; FLMNet-KD exceeds it on every benchmark on which it was judged. With three to six seeds, these are leads under a rule fixed in advance, not tests of statistical significance. The training cost stays close to that of QNN-$\alpha$. FLMNet-CF takes $46.7$\,ms per optimizer step against $45.4$ and $47.0$\,ms, because the two extra matrix multiplications per layer that the closed form needs (Proposition~\ref{prop:cheapvar}) roughly offset the second pass of the network body that it saves; its peak memory is about $50\%$ higher (Table~\ref{tab:exp-cost}). All models predict with a single forward pass.

\begin{table}[htbp]\centering\small

\caption{Training and inference cost on Criteo, measured on one Tesla V100 (32\,GB) in fp32 at batch size 4{,}096, with batches preloaded on the device. Forward passes count passes of the network body per optimizer step; QNN-$\alpha$ looks up the embeddings once and runs its body twice. Step time is the median over 300 timed steps after 50 warm-up steps, and includes each model's own training step (the EMA update for FLMNet-CF, the regularizer of FuxiCTR for QNN-$\alpha$). Peak memory is the peak allocated memory in MiB minus the preloaded batches. Inference latency is the median time in ms of one forward pass on a batch of 4{,}096 rows. The FLMNet-KD step excludes the teacher, whose logits are computed once in advance; its training runs count $K$ teachers and one student.}\label{tab:exp-cost}

\begin{tabular}{p{.21\linewidth}p{.14\linewidth}p{.11\linewidth}p{.12\linewidth}p{.13\linewidth}p{.13\linewidth}}

\toprule
Model & Forward passes / step & Step time (ms) & Peak memory (MiB) & Training runs & Inference latency (ms) \\ \midrule

QNN-$\alpha$ (one head) & 2 & 45.4 & 1106 & 1 & 5.9 \\

QNN-$\alpha$ (two heads) & 2 & 47.0 & 1108 & 1 & 5.7 \\

FLMNet-CF (ours) & 1 & 46.7 & 1667 & 1 & 5.8 \\

FLMNet-KD (ours) & 1 & 24.1 & 797 & $K+1$ & 5.8 \\

\bottomrule\end{tabular}\end{table}

Overall, a single FLMNet tower is the most accurate of the evaluated models on all five benchmarks, consistently across seeds. Its accuracy rests on three parts. Gates that read the clean pass make the closed-form moments exact by (C1), and their field-local form recovers nearly all of the accuracy that QNN-$\alpha$ loses when its gate reads the input embedding (Section~\ref{sec:exp-tower}). The closed-form penalty outperforms the sampling-based regularizers tested, still leads dropout at matched dose on Criteo and KKBox, and works mainly by slowing overfitting (Section~\ref{sec:exp-mask}). Cross-seed distillation then carries part of the seed-ensemble gain into the tower, largely on top of the gain from the penalty (Section~\ref{sec:exp-seed}).

\FloatBarrier

\section{Related Work}
\label{sec:related}

CTR prediction has largely been a search for the right way to form feature interactions. Logistic regression leaves every cross term to hand engineering~\cite{richardson2007}, factorization machines learn pairwise interactions as inner products of latent vectors~\cite{rendle2010}, and embedding-and-MLP models leave all interaction to depth~\cite{covington2016}. Later models combine explicit and implicit interactions, through a product layer~\cite{ipnn2016}, a parallel FM branch~\cite{deepfm2017}, or explicit layers that raise the interaction order step by step~\cite{xdeepfm2018,dcnv22021,fcn2026}, with bridges between parallel towers in EDCN~\cite{edcn2021}. Others make the interaction adaptive, through self-attention over fields~\cite{autoint2019}, learned exponents~\cite{afn2020} or complex-valued orders~\cite{eulernet2023}, or multiply representations by learned gates~\cite{masknet2021,finalmlp2023}. Quadratic towers make the product explicit, multiplying a linear projection elementwise with an activated one~\cite{final2023}, or forming a multi-head Khatri--Rao product between each head of the input and an activated projection of the whole input~\cite{qnn2025}. These designs are now compared under the open protocol of BARS~\cite{bars2022}.

A second line of work supervises a network with an ensemble. Distillation trains a network to match the outputs of a teacher~\cite{hinton2015}, born-again networks take as teacher an earlier network of the same architecture~\cite{bornagain2018}, and in CTR prediction, distilling an ensemble of independently trained models into one network yields a stronger single model~\cite{ensemblekd2020}. Its authors attribute the gain to teachers that differ in architecture or in the split of the training data. Towers of one architecture that differ only in their seed already form such a teacher (Section~\ref{sec:exp-seed}), and towers that differ only in the order of the training examples disagree almost as much as towers from different seeds and gain almost as much when pooled (Appendix~\ref{app:order-only}). The gain therefore need not come from distinct views. We trace it to the course of training: different initializations and data orders already accumulate substantial disagreement, and Appendix~\ref{app:order-only} re-reads their six-teacher result in this light. In concurrent work, Chopra et al. also distill the mean logit of independent runs into one CTR model and attribute the gain to the predictive variation that the runs do not share~\cite{recap2026}, which is the second-order form of the gap in \eqref{eq:moments}. Whether the views of other architectures add to this disagreement and raise AUC further remains open; answering it would separate what the training trajectory contributes from what a distinct view contributes. Self-ensemble methods build the teacher from the network itself: R-Drop penalizes the divergence between two dropout samples~\cite{rdrop2021}, FinalNet supervises each of two parallel blocks with their averaged prediction~\cite{final2023}, and QNN averages two dropout passes into a frozen teacher~\cite{qnn2025}. Such teachers are built from a few sampled masks and so carry sampling error. Analytic treatments of dropout remove the sampling instead, through a Gaussian approximation~\cite{fastdropout2013} or an expansion of the expected objective~\cite{wager2013,wei2020}. Our work builds on these lines: it trains a gated quadratic tower with the closed-form limit of the self-ensemble objective, and distills an ensemble of its seeds into one tower.

\section{Discussion and Conclusion}

FinalNet and QNN report gains in CTR prediction from self-ensemble objectives, but do not explain what the added term regularizes. We answered this for members pooled in logit space. As the number of masks grows without bound, the masked self-ensemble objective converges exactly to the label loss at the mean logit plus a multiple of the Jensen gap of the softplus over masks, so the limit regularizes only the disagreement among masked members, which in practice mainly slows overfitting. When the gates of a tower do not read the masked path, as the field-local gates of FLMNet ensure, the moments of the logit have closed forms, and a Gaussian surrogate of the gap trains the limit in one deterministic forward pass. Independently trained seeds give rise to a gap of the same form, which has no closed form; we reduce it by distilling an ensemble of seeds into the tower. On all five benchmarks, the resulting single tower is the most accurate of the evaluated models, consistently across seeds, and predicts in one forward pass. The gain of the closed-form penalty largely adds to that of cross-seed distillation.

These results split the advantage of an ensemble over a single tower into two sources. The first is mask noise, the disagreement among masks. This is where the gain of self-ensemble distillation comes from, and the closed-form penalty reduces it during training without sampled forward passes. The second is seed noise, the disagreement among seeds. It accounts for the further gain of an ensemble, and cross-seed distillation reduces it after training without changing the architecture. FLMNet-CF collects the first gain at a per-step training cost close to that of QNN-$\alpha$, and FLMNet-KD the second at the price of $K+1$ training runs; conditions (C1)--(C3) (Section~\ref{sec:tower}) give a concrete criterion for designing towers that admit the closed form.

These conclusions have three limits. The closed form is exact only for towers that satisfy (C1)--(C3). Training replaces the exact gap by a Gaussian surrogate. The gap across seeds has no closed form, so how much of it distillation recovers is known only from measurement. The two gains are also collected separately: the penalty within one training run and distillation over $K+1$ runs. Two directions merit future work. The first is to collect both gains in a single training run, for example by training several towers jointly and serving them in one forward pass. The second is to adapt existing CTR architectures to satisfy (C1)--(C3) and then distill them across seeds, which may raise their AUC as well.

\begin{acks}
We thank B. Zhu and C. Cai for inspiring discussions.
\end{acks}

\appendix
\section{Proofs and Derivations}
\subsection{Why Higher-Order Truncation Need Not Help}
\label{app:orders}
Two properties of the softplus limit the moment expansion \eqref{eq:moments}. First, $\varsigma(z)=\log(1+e^z)$ is singular at $z=\pm i\pi$, so its Taylor series at $\mu$ converges only for $|\delta|<\sqrt{\mu^2+\pi^2}$. For masks that move the logit further, the expansion does not converge, and adding a term can increase the error. Second, the coefficients are derivatives at $\mu$ alone. When $\sigma(\mu)$ is saturated they are small, while a masked logit may reach the region where $\varsigma$ bends: at $\mu=-5$ and $\delta=5$, the second-order term gives $0.083$ against an exact contribution $\varsigma(\mu+\delta)-\varsigma(\mu)-\sigma(\mu)\delta=0.653$. Both concern masked logits far from $\mu$, and Figure~\ref{fig:surrogate} shows the truncations falling short on rows with large $V$. The Gaussian proxy avoids the series altogether and evaluates $\varsigma$ away from $\mu$, at the nodes $\mu+\xi_i\sqrt V$.

\subsection{Dose: \texorpdfstring{$\kappa$}{kappa} and \texorpdfstring{$\rho$}{rho} to Leading Order}
\label{app:kappa-rho}
In \eqref{eq:hoeffding} each $\partial_Tz$ is a mixed derivative of the multilinear $z_m$ evaluated at the all-ones mask, a function of $\theta$ and $X$ that does not involve $\rho$; the masking rate enters only through the weights $\omega^{|T|}$. Hence $V=\omega V_1+O(\omega^2)$ with $V_1=\sum_{|T|=1}(\partial_Tz)^2$, and since $V-\widehat V$ consists of terms of order $\omega^2$ (Proposition~\ref{prop:cheapvar}), also $\widehat V=\omega V_1+O(\omega^2)$, which is the quantity the penalty uses. Three-point Gauss--Hermite quadrature is exact for polynomials of degree five, so the penalty of \eqref{eq:trained}, evaluated at $\widehat V$, expands as $\tfrac12\varsigma''(\mu)\widehat V+\tfrac18\varsigma''''(\mu)\widehat V^2+O(\widehat V^3)$. Together,
\[
\kappa\Big[\sum_{i}\lambda_i\,\varsigma\big(\mu+\xi_i\sqrt{\widehat V}\big)-\varsigma(\mu)\Big]=\tfrac12\,\kappa\omega\,\varsigma''(\mu)V_1+O(\kappa\omega^2),
\]
so to first order $\kappa$ and $\rho$ act only through the dose $\kappa\omega=\kappa\rho/(1-\rho)$. The product law fails in two places, both of order $\omega^2$: the terms with $|T|\ge2$, which make $\widehat V$ superlinear in $\omega$, and the curvature of the proxy in $\widehat V$, which bends it away from linear once $\widehat V$ is not small.

\subsection{Computing \texorpdfstring{$\widehat V$}{V-hat} from the Clean Pass}
\label{app:moments}
Given the quantities cached by the clean forward pass, $\widehat V$ of Proposition~\ref{prop:cheapvar} is evaluated per example as follows, with $F_l$ absorbed as in Appendix~\ref{app:absorb}.
(i) For each layer, $K_{l,jj}=\sum_tU_{l,jt}^2r_{l,t}^2$, one product of $U_l^{\odot2}$ with $r_l^{\odot2}$.
(ii) $V_{\mathrm{diag}}=\sum_ja_j^2D_{L,j}$ by the recursion \eqref{eq:vdiag-rec}, from $D_0=0$ and $\Pi_0=\mathbf 1$ with $\Pi_l=\Pi_{l-1}\odot(1+g_l)^{\odot2}$.
(iii) The prefix products $\prod_{l'<l}(1+g_{l'})$ and suffix products $\prod_{l'>l}(1+g_{l'})$, whose product is $P_{-l}$.
(iv) For each layer, $t_l=U_l^{\top}(a\odot P_{-l})$, the second product, and $V_{\mathrm{off}}$ accumulates $\omega\big[\|r_l\odot t_l\|^2-\sum_ja_j^2P_{-l,j}^2K_{l,jj}\big]$.
Steps (i) and (iv) are the two matrix products per layer of Section~\ref{sec:closed-form}; everything else is elementwise. The computation uses only the independence of all mask coordinates, within and across layers, as in Theorem~\ref{thm:exactvar}.

\subsection{Exactness of the Recalibration Absorption}
\label{app:absorb}
When $D^{\mathrm{fg}}\ge2$, the $i$-th recalibration multiplies the input of layer $i$, that is the output of layer $i-1$, by a vector that depends on $X$ alone. Write $F_l\in\mathbb R^{h}$ for the factor multiplying the output of layer $l$, with $F_l=\mathbf 1$ at the layers where no recalibration applies, so that the carrier of \eqref{eq:tower} becomes $x_l=F_l\odot x_{l-1}\odot(1+g_l)$ and the masked logit becomes, in the notation of \eqref{eq:zm-product},
\[
z_m=b_0+\sum_ja_j\prod_{l=1}^{L}F_{l,j}\,\pi_{l,j},\qquad
\pi_{l,j}=v_{l,j}\big(1+g_{l,j}+\eta_{l,j}\big),\quad\eta_l=U_l\big((u_l-1)\odot r_l\big).
\]
Define the relabelled layer parameters
\[
1+g'_{l,j}=F_{l,j}\big(1+g_{l,j}\big),\qquad U'_l=\mathrm{diag}(F_l)\,U_l .
\]
Then $\eta'_l=U'_l\big((u_l-1)\odot r_l\big)=F_l\odot\eta_l$, and for every layer, coordinate and mask
\begin{equation}
\label{eq:absorb}
v_{l,j}\big(1+g'_{l,j}+\eta'_{l,j}\big)
=v_{l,j}F_{l,j}\big(1+g_{l,j}+\eta_{l,j}\big)
=F_{l,j}\,\pi_{l,j}.
\end{equation}
The recalibrated network is therefore the unrecalibrated one with $(g_l,U_l)$ replaced by $(g'_l,U'_l)$, as a pointwise identity in $m$ and not merely in distribution. Proposition~\ref{prop:mean}, Theorem~\ref{thm:exactvar} and Proposition~\ref{prop:cheapvar} accordingly hold verbatim for the primed quantities, which is what the statement of Section~\ref{sec:closed-form} means by absorbing $F_l$ into the layer. Concretely, $1+g_{l,j}$ is replaced by $F_{l,j}(1+g_{l,j})$ throughout, $K_{l,jj'}$ by $F_{l,j}F_{l,j'}K_{l,jj'}$ in \eqref{eq:KQ}, and $U_l^{\top}(a\odot P_{-l})$ by $U_l^{\top}(F_l\odot a\odot P_{-l})$ in \eqref{eq:voff}.

Two remarks. First, $r_l$ is untouched by the substitution, because the gate reads $x_0$ rather than the carrier; this is (C1) doing the work, and had the gate read $x_{l-1}$ the recalibration would also move $r_l$ and \eqref{eq:absorb} would fail. Second, both replacements are needed. Applying only $1+g_l\leftarrow F_l\odot(1+g_l)$ reproduces the forward pass, the mean $\mu$ and the parameter count exactly while leaving $\widehat V$ silently biased, so the usual sanity checks do not catch the omission.

\subsection{Proof of Proposition~\texorpdfstring{\ref{prop:cheapvar}}{4.3}}
\label{app:cheapvar}
Throughout, $Q_{l,jj'}$ is as in \eqref{eq:KQ} and $U_l$ already absorbs the recalibration factor $F_l$.

\emph{Diagonal terms.} For $j=j'$ equation \eqref{eq:KQ} reads $Q_{l,jj}=(1+\omega)\big[(1+g_{l,j})^2+\omega K_{l,jj}\big]$, and substituting it into \eqref{eq:exactvar} gives \eqref{eq:vdiag} directly; no truncation is involved, so $V_{\mathrm{diag}}$ is exact in $\omega$. For the recursion, put
\[
B_{l,j}=Q_{l,jj},\qquad C_{l,j}=(1+g_{l,j})^2,\qquad
\Pi_{l,j}=\prod_{l'\le l}C_{l',j},\qquad
D_{l,j}=\prod_{l'\le l}B_{l',j}-\Pi_{l,j},
\]
so that $D_{0,j}=0$ and $V_{\mathrm{diag}}=\sum_ja_j^2D_{L,j}$. Then
\[
D_{l}=B_{l}\Big(D_{l-1}+\Pi_{l-1}\Big)-C_{l}\Pi_{l-1}=B_{l}D_{l-1}+\Pi_{l-1}\big(B_{l}-C_{l}\big),
\]
which is \eqref{eq:vdiag-rec}; expanding $B_{l,j}-C_{l,j}$ and cancelling the common $(1+g_{l,j})^2$ gives the increment stated there. Since $K_{l,jj}=\sum_t U_{l,jt}^2r_{l,t}^2\ge0$ and $\omega>0$, the three quantities $B_l$, $D_{l-1}$ and $\Pi_{l-1}(B_l-C_l)$ are nonnegative, so nonnegativity of $D_{L,j}$ follows by induction without any cancellation of large terms.

\emph{Off-diagonal terms.} For $j\neq j'$ the Kronecker prefactor is one, so with $c_{l}=(1+g_{l,j})(1+g_{l,j'})$ we have $Q_{l,jj'}=c_l+\omega K_{l,jj'}$ and
\begin{align*}
\prod_{l}Q_{l,jj'}-\prod_{l}c_{l}
&=\omega\sum_{l}K_{l,jj'}\prod_{l'\neq l}c_{l'}+O(\omega^2)\\
&=\omega\sum_{l}K_{l,jj'}P_{-l,j}P_{-l,j'}+O(\omega^2).
\end{align*}
Keeping the first order and writing $s^{(l)}=a\odot P_{-l}$, the off-diagonal contribution to \eqref{eq:exactvar} is $\omega\sum_lT_l$, where $T_l$ collects the pairs of layer $l$. Completing the inner sum over all pairs and subtracting the diagonal,
\[
T_l=\sum_{j\neq j'}s^{(l)}_js^{(l)}_{j'}K_{l,jj'}
=\big(s^{(l)}\big)^{\!\top}K_ls^{(l)}-\sum_j\big(s^{(l)}_j\big)^2K_{l,jj}
=\big\|r_l\odot U_l^{\top}s^{(l)}\big\|^2-\sum_j\big(s^{(l)}_j\big)^2K_{l,jj},
\]
the last equality because $K_l=U_l\mathrm{diag}(r_l^{\odot2})U_l^{\top}$ implies $s^{\top}K_ls=\big\|\mathrm{diag}(r_l)U_l^{\top}s\big\|^2$. This is \eqref{eq:voff}. The only terms dropped are those with $j\neq j'$ carrying two or more factors of $\omega$, which are the cross-layer off-diagonal remainders. \qed

\section{Experimental Details}
\label{app:repro}

\subsection{Architecture and Penalty}
\label{app:arch}

Table~\ref{tab:app-arch} lists the structure of FLMNet-CF on each benchmark. The tower width is $h=Cd$, fixed by the number of fields $C$ and the embedding dimension $d$; $L$ is the number of layers and $M$ the number of gate units per tower coordinate (Section~\ref{sec:tower}). Every benchmark uses the masking rate $\rho=0.1$, the closed-form penalty with the coefficient $\kappa$ in the table, field-local gates with recalibration at the first $D^{\mathrm{fg}}=2$ layers, and ReLU gate units. The embedding tables are initialized from a zero-mean normal distribution with the standard deviation in the table.\par

\begin{table}[htbp]\centering\small
\caption{Structure of FLMNet-CF on each benchmark. Parameter counts are for the tower and for the embedding tables, in millions.}\label{tab:app-arch}
\begin{tabular}{lrrrrrrrrr}
\toprule
\textbf{Benchmark} & $d$ & $C$ & $h$ & $L$ & $M$ & $\kappa$ & \textbf{Emb.\ init std} & \textbf{Tower} & \textbf{Emb.} \\ \midrule
Criteo & 10 & 39 & 390 & 4 & 5 & 2 & $10^{-4}$ & 3.14 & 20.87 \\
Avazu & 10 & 22 & 220 & 3 & 5 & 0.25 & $10^{-4}$ & 0.77 & 12.99 \\
KKBox & 128 & 13 & 1664 & 2 & 5 & 3 & 0.6 & 29.84 & 11.74 \\
Frappe & 10 & 10 & 100 & 2 & 13 & 1 & $10^{-4}$ & 0.29 & 0.05 \\
ML-1M & 16 & 7 & 112 & 2 & 13 & 1 & $10^{-4}$ & 0.38 & 0.16 \\
\bottomrule\end{tabular}
\end{table}

\subsection{Avazu Re-split}
\label{app:avazu}

The BARS documentation of Avazu describes a random 7:1:2 split, but the released validation and test splits behave like time blocks. Their positive rates are 0.1679 and 0.1525, against 0.1750 in the training split, and the gap between the two is about 69 standard errors. Comparing models on such a test split confounds model quality with the shift between the two days. We therefore keep the training split of 28,300,276 rows unchanged and pool the released validation split (4,042,897 rows) and test split (8,085,794 rows) into 12,128,691 rows. A random permutation with the fixed seed 20260929 assigns the first 4,042,897 rows to the new validation split and the remaining 8,085,794 rows to the new test split. The new splits have positive rates 0.157599 and 0.157625, and the two-proportion test for their difference gives $z=-0.12$. The training split is untouched, so the vocabularies, the number of steps per epoch and the schedule budgets carry over from the original board. The new validation and test splits still differ from the training split in positive rate by about 0.0174. This extrapolation in time is a property of the released training split and is kept by design.\par

One fact limits what the re-split can guarantee, and we state it here. The released validation split was used for model selection before the re-split. Two thirds of its rows (2,695,266) fall into the new test split, where they make up 33.33\% of the rows, and the remaining 1,347,631 fall into the new validation split. To rule out a selection that leans on these rows, every model is retrained, with early stopping and model selection on the new validation split, and the configuration of our tower was scanned again on it. The rescan found no configuration that improved on the one selected on the released split by more than the paired-difference scale of $3\times10^{-4}$ AUC, so the delivered configuration is unchanged and its selection does not rely on any test data. Every reading reported in the main tables comes from the new test split, which was revealed once under the protocol of Appendix~\ref{app:protocol}. For completeness, Table~\ref{tab:app-avazu-x1} reports every model on the released split, with the same configurations and the checkpoints selected on the released validation split.\par

\begin{table}[htbp]\centering\small
\caption{Test results on the released Avazu split, before the re-split. Entries are means over the seeds in the last column; standard deviations of test AUC are over the same seeds.}\label{tab:app-avazu-x1}
\setlength{\tabcolsep}{6pt}
\begin{tabular}{lcccc}
\toprule
\textbf{Model} & \textbf{Logloss} & \textbf{AUC} & \textbf{AUC s.d.} & \textbf{Seeds} \\ \midrule
LR & 0.3736 & 0.7514 & $1.5\times10^{-4}$ & 3 \\
FM & 0.3683 & 0.7602 & $5.8\times10^{-4}$ & 3 \\
DNN & 0.3677 & 0.7629 & $4.3\times10^{-4}$ & 3 \\
IPNN & 0.3680 & 0.7618 & $1.9\times10^{-4}$ & 3 \\
DeepFM & 0.3679 & 0.7625 & $1.8\times10^{-3}$ & 3 \\
xDeepFM & 0.3674 & 0.7639 & $2.0\times10^{-4}$ & 3 \\
AutoInt+ & 0.3675 & 0.7640 & $4.3\times10^{-4}$ & 3 \\
AFN+ & 0.3675 & 0.7642 & $6.3\times10^{-4}$ & 3 \\
DCNv2 & 0.3684 & 0.7624 & $5.7\times10^{-4}$ & 3 \\
EDCN & 0.3676 & 0.7634 & $1.4\times10^{-3}$ & 6 \\
MaskNet & 0.3692 & 0.7604 & $1.6\times10^{-4}$ & 3 \\
FinalMLP & 0.3658 & 0.7660 & $7.0\times10^{-4}$ & 6 \\
EulerNet & 0.3676 & 0.7633 & $5.1\times10^{-4}$ & 3 \\
FinalNet (2B) & 0.3658 & 0.7664 & $3.6\times10^{-4}$ & 6 \\
QNN-$\alpha$ (one head) & 0.3698 & 0.7611 & $2.0\times10^{-3}$ & 6 \\
QNN-$\alpha$ (two heads) & 0.3687 & 0.7610 & $1.4\times10^{-3}$ & 6 \\
ECN & 0.3688 & 0.7602 & $7.6\times10^{-4}$ & 6 \\
FLMNet-CF (ours) & 0.3657 & 0.7661 & $3.3\times10^{-4}$ & 6 \\
FLMNet-KD (ours) & 0.3657 & 0.7660 & $2.5\times10^{-4}$ & 3 \\
\bottomrule\end{tabular}
\end{table}

\subsection{Training}
\label{app:train}

\paragraph{Per-benchmark training settings.} Table~\ref{tab:app-train} lists the settings of FLMNet-CF on each benchmark. Every FLMNet-CF run uses Adam, gradient clipping at 10, single precision, and weight decay on the embedding tables only, and the delivered weights are the exponential moving average (EMA) of the training weights. Where a schedule is given, the learning rate follows a cosine decay from its peak value to $10^{-5}$, and the checkpoint of the best validation epoch is delivered. On Criteo and Avazu the cosine schedule has a fixed epoch budget together with early stopping on validation AUC (patience 2). KKBox holds the peak rate for 22 epochs and decays it over the remaining 8, without early stopping. Frappe and ML-1M keep the stopping rule of FuxiCTR, which multiplies the learning rate by 0.1 (floor $10^{-6}$) after every epoch without validation improvement and stops after two consecutive such epochs; we use it as an explicit legacy protocol on these two boards, and the learning rate in Table~\ref{tab:app-train} is then the initial value. Batch size, peak learning rate, weight decay and the averaging horizon were chosen on the validation split of each benchmark.\par

\begin{table}[htbp]\centering\small
\caption{Training settings of FLMNet-CF. The EMA decay $\beta$ is applied per optimizer step. On KKBox and ML-1M it corresponds to a horizon of 2.5 and 3 epochs, respectively. The FuxiCTR rule is described in the text; the two boards that use it are marked as a legacy protocol.}\label{tab:app-train}
\setlength{\tabcolsep}{4pt}
\begin{tabular}{@{}lccccp{.36\linewidth}@{}}
\toprule
\textbf{Benchmark} & \begin{tabular}[b]{@{}c@{}}\textbf{Batch}\\\textbf{size}\end{tabular} & \begin{tabular}[b]{@{}c@{}}\textbf{Peak}\\\textbf{LR}\end{tabular} & \begin{tabular}[b]{@{}c@{}}\textbf{Emb. weight}\\\textbf{decay}\end{tabular} & \textbf{EMA $\beta$} & \textbf{LR schedule and stopping rule} \\ \midrule
Criteo & 4096 & $10^{-3}$ & $10^{-5}$ & 0.9999 & Cosine over 40 epochs, early stopping (patience 2) \\
Avazu & 4096 & $5\times10^{-4}$ & $2.5\times10^{-3}$ & 0.9999 & Cosine over 20 epochs, early stopping (patience 2) \\
KKBox & 10000 & $10^{-3}$ & $3\times10^{-4}$ & 0.9993178 & Constant for 22 epochs, cosine over the last 8 of 30, no early stopping \\
Frappe & 1024 & $8\times10^{-3}$ & $10^{-3}$ & 0.99949088 & FuxiCTR rule (legacy) \\
ML-1M & 1024 & $2\times10^{-3}$ & $5\times10^{-4}$ & 0.99942330 & FuxiCTR rule (legacy) \\
\bottomrule\end{tabular}
\end{table}

\paragraph{Cross-seed distillation.} Every FLMNet-KD student starts from the delivered EMA weights of one FLMNet-CF tower and regresses a fixed teacher, the arithmetic mean of the logits of $K=3$ FLMNet-CF towers that include its starting tower, computed once before the student is trained (Eq.~\eqref{eq:kd}). The student sees no label and no penalty, uses a fresh Adam optimizer at a learning rate of $10^{-4}$ with the batch size and embedding weight decay of FLMNet-CF on the same benchmark, and stops under the FuxiCTR rule on validation AUC, with at most 60 epochs after it starts; it is evaluated with its raw weights. On Criteo, each of six students, one per seed 110000--110005, has its own teacher, which pools its starting tower with two of the other five towers drawn at random once and then fixed. On the other benchmarks, three students start from three towers and their teachers pool those same three: seeds 110000--110002 on Avazu and KKBox, 110008--110010 on Frappe, and 110002--110004 on ML-1M.\par

\paragraph{Seed counts and shared members.} Table~\ref{tab:app-seeds} collects the numbers of random seeds stated in the captions of Tables~\ref{tab:exp-main} and~\ref{tab:exp-main-b}. The main competitors are EDCN, FinalMLP, FinalNet, QNN-$\alpha$ and ECN. On ML-1M, eight FLMNet-CF seeds are reported. The students of FLMNet-KD on each benchmark distil from teachers built on overlapping sets of FLMNet-CF towers, so they are not independent repetitions, and the counts for FLMNet-KD in Table~\ref{tab:app-seeds} should be read in that sense. On ML-1M, for example, the three students share a single teacher ensemble.\par

\begin{table}[htbp]\centering\small
\caption{Number of random seeds behind each entry of Tables~\ref{tab:exp-main} and~\ref{tab:exp-main-b}. FLMNet-KD students on a benchmark are built on overlapping sets of FLMNet-CF towers and are not independent.}\label{tab:app-seeds}
\setlength{\tabcolsep}{6pt}
\begin{tabular}{lcccc}
\toprule
\textbf{Benchmark} & \textbf{Main competitors} & \textbf{FLMNet-CF} & \textbf{FLMNet-KD} & \textbf{Remaining baselines} \\ \midrule
Criteo & 6 & 6 & 6 & 3 \\
Avazu & 6 & 6 & 3 & 3 \\
KKBox & 6 & 6 & 3 & 3 \\
Frappe & 6 & 6 & 3 & 3 \\
ML-1M & 6 & 8 & 3 & 3 \\
\bottomrule\end{tabular}
\end{table}

\subsection{Baselines}
\label{app:baselines}

\paragraph{Baseline configurations.} Each baseline is trained with the best configuration recorded publicly for it: the BARS configuration where BARS lists the model, and otherwise the configuration in the authors' released log or the hyper-parameters of the closest BARS configuration, as stated in the notes to Tables~\ref{tab:exp-main} and~\ref{tab:exp-main-b}. Stopping follows each original configuration, model by model, and apart from the batch size of QNN-$\alpha$ on Criteo, noted below, the baselines are not retuned on our validation splits. Only board-level quantities are aligned across models. The embedding dimension is 10 on Criteo, Avazu and Frappe, 128 on KKBox (16 for EulerNet) and 16 on ML-1M. Where the original configuration of a model used another value, only the affected keys were changed. On Criteo, EulerNet moves from dimension 16 to 10 and from batch size 10000 to 4096, and QNN-$\alpha$, with one and with two heads, moves from the batch size 10000 of the authors' log to 4096, the standard batch size of BARS, after a check on the Criteo validation split favoured 4096 (AUC 0.81505 against 0.81424 with one head). On Avazu and Frappe, ECN borrows its Criteo configuration and moves from batch size 8192 to 4096. On ML-1M, ECN keeps its KKBox configuration except for dimension 16 and batch size 4096, and EulerNet keeps it except for batch size 4096; the other baselines already use batch size 4096 in their BARS \texttt{movielenslatest\_x1} configurations, so their only board-level change is the dimension, from 10 to 16. On the re-split Avazu, the original configurations of all seventeen baselines are retrained on the new training and validation splits, 69 runs in total.\par

\paragraph{The QNN authors' configuration on ML-1M.}\label{app:ml1m-qnn4} Tables~\ref{tab:exp-main} and~\ref{tab:exp-main-b} report QNN-$\alpha$ with one and with two heads on every benchmark. For ML-1M the QNN authors also release a training log of the configuration they selected, which uses four heads. We retrained that configuration over six seeds with every hyper-parameter copied from the log; it is the strongest baseline we trained on this board, and we report it here only because the main tables have no room for a row that exists on a single benchmark. It reaches a test Logloss of 0.3028 and a test AUC of 0.9058. Over the six seeds it shares with FLMNet-CF, FLMNet-CF is higher in AUC by $3.0\times10^{-3}$ on average and on every seed, by no less than $1.8\times10^{-3}$; over the three seeds of FLMNet-KD, FLMNet-KD is higher by $3.9\times10^{-3}$ on average and by no less than $3.3\times10^{-3}$.\par

\subsection{Reproduction Gaps}
\label{app:gaps}

\paragraph{Published values on Frappe.} On Frappe our seed-averaged reproductions of FinalNet, FinalMLP and EDCN fall below the test AUC that BARS publishes for them, by $1.1\times10^{-3}$, $1.2\times10^{-3}$ and $9.8\times10^{-4}$. The published values are not comparable to seed averages. For each of these models BARS runs a grid of hyper-parameter configurations and publishes the configuration with the highest test AUC, which is not the one with the highest validation AUC: the published FinalNet configuration ranks second of 16 on validation AUC, and the configuration ranked first on validation reaches a test AUC of 0.98554 rather than the published 0.98574; the published FinalMLP and EDCN configurations likewise rank second of 32 and of 28, and MaskNet third of 32. Our configurations match the published ones in every hyper-parameter except the random seed, and no configuration key is silently dropped by FuxiCTR~2.3.9. EulerNet, whose published Frappe value is a single run from the authors' log, falls $1.7\times10^{-3}$ below it; rerunning the authors' configuration with their seed in our pipeline reproduces their validation AUC to within $1.8\times10^{-4}$, and our three seeds spread with a validation-AUC standard deviation of $9.4\times10^{-4}$, so the published run sits at the favourable end of that spread.\par

\paragraph{Reproduction gaps on the other benchmarks.} The differences between our seed-averaged reproductions and published values are not confined to Frappe. On Criteo, 11 of the 13 baselines that BARS lists fall within $3\times10^{-4}$ of the published validation AUC (of the published range, where BARS reports several runs); the two exceptions are FinalNet and FinalMLP, which are the models for which BARS reports a single run, and their gaps on the test split are given in the caption of Table~\ref{tab:exp-main}. On KKBox, we compare 13 baselines with the test AUC declared for them in BARS, in the QNN authors' log or in a published table. Six agree within $3\times10^{-4}$. Six fall below the declared value, by $4.8\times10^{-4}$ to $9.5\times10^{-4}$ (FM, IPNN, xDeepFM, AutoInt+, AFN+ and QNN-$\alpha$ with one head), and MaskNet lies above it by $1.3\times10^{-3}$. On every one of these models the gap has the same sign on the validation split. As on Frappe, the declared values are single runs, while ours are averages over several seeds. On the released Avazu split we compare 14 baselines in the same way (Table~\ref{tab:app-avazu-repro}). Avazu after the re-split and ML-1M have no declared values to compare with.

\medskip\noindent\textit{Released Avazu split.} We compare the test AUC of 14 baselines with the value declared for them in BARS or, for EulerNet, in the authors' log. QNN-$\alpha$ (one and two heads) and ECN have no declared value. Eight baselines agree with the declared value within a tolerance of $2\,\mathrm{sd}\sqrt{1/n_{\mathrm{decl}}+1/n_{\mathrm{ours}}}$, where sd is the standard deviation of the test AUC over the repeated BARS runs (for models with a single declared run, the median of the four models that have five), and $n_{\mathrm{decl}}$ and $n_{\mathrm{ours}}$ are the numbers of declared runs and of our seeds. This tolerance ranges from $5.4\times10^{-4}$ to $2.5\times10^{-3}$ across models. Six baselines fall below the declared value, by $1.0\times10^{-3}$ to $3.9\times10^{-3}$ (FM, IPNN, DeepFM, EDCN, MaskNet and EulerNet), and none lies above it. Thirteen of the 14 gaps are negative. Only DeepFM, whose declared value is a five-run mean, is also below the declared value on the validation split; IPNN, EDCN, MaskNet and EulerNet agree with it there, and FM lies above it by $6.7\times10^{-3}$.

\begin{table}[htbp]\centering\small
\caption{Test AUC of our reproductions on the released Avazu split against the declared value. Ours is the mean over seeds; gap is ours minus declared, in units of $10^{-4}$. The tolerance is defined in the text. The last column gives the reading of the gap.}\label{tab:app-avazu-repro}
\setlength{\tabcolsep}{5pt}
\begin{tabular}{lccrll}
\toprule
\textbf{Model} & \textbf{Declared} & \textbf{Ours} & \textbf{Gap} & \textbf{Source} & \textbf{Reading} \\ \midrule
LR & 0.751625 & 0.7513945 & $-2.305$ & BARS, 1 run & within tolerance \\
FM & 0.761253 & 0.7602211 & $-10.32$ & BARS, 1 run & below \\
DNN & 0.763827 & 0.7628807 & $-9.463$ & BARS, 1 run & within tolerance \\
IPNN & 0.763014 & 0.7617708 & $-12.43$ & BARS, 1 run & below \\
DeepFM & 0.764586 & 0.7624552 & $-21.31$ & BARS, mean of 5 runs & below \\
xDeepFM & 0.764205 & 0.7639083 & $-2.967$ & BARS, mean of 5 runs & within tolerance \\
AutoInt+ & 0.764524 & 0.7640333 & $-4.907$ & BARS, 1 run & within tolerance \\
AFN+ & 0.764145 & 0.7642193 & $+0.7430$ & BARS, mean of 5 runs & within tolerance \\
DCNv2 & 0.763639 & 0.7624330 & $-12.06$ & BARS, mean of 5 runs & within tolerance \\
EDCN & 0.765241 & 0.7634479 & $-17.93$ & BARS, 1 run & below \\
MaskNet & 0.764289 & 0.7603693 & $-39.20$ & BARS, 1 run & below \\
FinalMLP & 0.766566 & 0.7660436 & $-5.224$ & BARS, 1 run & within tolerance \\
FinalNet & 0.767159 & 0.7664078 & $-7.512$ & BARS, 1 run & within tolerance \\
EulerNet & 0.764515 & 0.7632749 & $-12.40$ & Authors' log, 1 run & below \\
\bottomrule\end{tabular}
\end{table}

Ten of the 14 declared values are single runs (nine from BARS and the EulerNet author log), while ours are means over three or six seeds. A single run can lie on either side of the mean of its own configuration, and the tolerance above allows for this only through the borrowed standard deviation. For EDCN, the FuxiCTR version that we use has no switch for the regulation module, so it is always on in our runs, which is not the structure that BARS reports; whether this accounts for the gap of $1.8\times10^{-3}$ was not tested. For FM, our configuration matches the best BARS configuration in regularization strength and batch size; the BARS run used FuxiCTR 1.1.0 and ours uses 2.3.9, and the two builds of FM differ by 242 parameters, but whether this accounts for the gap was not tested. For IPNN, MaskNet and EulerNet, whose declared values are single runs and whose validation AUC agrees with the declared one, we found no difference in the main training hyper-parameters and have not identified the cause. For DeepFM, the declared value is a mean of five runs, so a single-run explanation does not apply, and we have not identified the cause either. Across the six models, the difference between test and validation AUC of our runs is below the range of $1.6\times10^{-2}$ to $2.2\times10^{-2}$ observed for the declared runs for 16 of our 21 runs (FM, IPNN, MaskNet and EulerNet on three of three seeds, EDCN on three of six, DeepFM on one of three). We have not identified the reason for this shift.\par

\subsection{Evaluation Protocol}
\label{app:protocol}

\paragraph{Test reveal protocol.} Model selection uses the validation split only. A test split is read only through a roster that is committed before any test reading, which lists every checkpoint to be reported by tag, seed and file checksum (MD5), together with the decision rule: the comparison, its threshold, and how an outcome in the opposite direction will be read. The reveal tool first recomputes the validation AUC of each listed checkpoint and requires agreement with the value recorded at training time (tolerance $2\times10^{-6}$), so a substituted or altered checkpoint is rejected before any test data is touched. It then writes a receipt for each member before computing the test metrics, refuses to overwrite an existing receipt, and refuses to rerun after an interruption. All members of the roster are reported and none is dropped or added. The once-only guarantee holds within one fixed output directory; the tool cannot see a reveal made into a different directory, so it rests on the procedure and the committed roster as well as on the tool. After a reveal, the configuration is frozen, and a later change would need a new roster and would have to be reported together with the order in which the two were run. Our towers were trained with the test data unread (blind mode), and on the re-split Avazu the baseline harness likewise never constructed a test iterator.\par

\paragraph{Seed-wise test judgements.} Table~\ref{tab:app-judge} lists, for each benchmark, the judgement fixed before the test reveal. The opponent is the strongest baseline on validation. For seed $s$, $\Delta_s$ is the test AUC of our model minus that of the opponent under the same seed index, and the rule compares the mean $\bar\Delta$ over $n$ seeds with the threshold $\theta=\max\bigl(F,\,2\,\mathrm{sd}(\Delta_s)/\sqrt n\bigr)$: $\bar\Delta\ge\theta$ is read as a lead, $\bar\Delta\le-\theta$ as no lead, and anything in between as no reading. The floor $F$ was fixed before the reveal. On Criteo and KKBox it is a paired threshold measured on an earlier Criteo board, $2.94\times10^{-5}$; on Avazu it is the same paired statistic computed on the validation split; on Frappe and ML-1M it is $3\times10^{-4}$. Paired seeds share only their index, not their randomness. On Frappe the rule is unpaired: $\bar\Delta$ is the difference of the two seed means, and the second term of $\theta$ is $2\sqrt{s_A^2/n_A+s_B^2/n_B}$. FLMNet-KD is judged against the same opponent on Criteo, Avazu, KKBox and ML-1M; on Frappe its rule reports only the mean and range of its students. On KKBox the judgement of FLMNet-CF uses five seeds, because the checkpoint of seed 110003 was not available when the roster was committed; that seed was revealed later under a supplementary roster and enters the main tables but not this judgement.\par

\begin{table}[htbp]\centering\small
\caption{Seed-wise test judgements. $\bar\Delta$, $F$ and $\theta$ are in units of $10^{-3}$ AUC; the rule and the floor $F$ are defined in the text. Sign: number of seeds with $\Delta_s>0$. On Frappe the rule is unpaired and FLMNet-KD is not judged.}\label{tab:app-judge}
\begin{tabular}{llllrrrrc}
\toprule
\textbf{Benchmark} & \textbf{Model} & \textbf{Opponent} & $n$ & $F$ & $\bar\Delta$ & $\theta$ & $\bar\Delta/\theta$ & \textbf{Sign} \\ \midrule
Criteo & CF & QNN-$\alpha$ (two heads) & 6 & 0.029 & 0.605 & 0.142 & 4.2 & 6/6 \\
 & KD & QNN-$\alpha$ (two heads) & 6 & 0.029 & 1.311 & 0.140 & 9.4 & 6/6 \\
Avazu & CF & FinalNet & 6 & 0.547 & 1.029 & 0.547 & 1.9 & 6/6 \\
 & KD & FinalNet & 3 & 0.758 & 1.325 & 0.758 & 1.7 & 3/3 \\
KKBox & CF & QNN-$\alpha$ (two heads) & 5 & 0.029 & 0.605 & 0.070 & 8.7 & 5/5 \\
 & KD & QNN-$\alpha$ (two heads) & 3 & 0.029 & 2.919 & 0.165 & 17.7 & 3/3 \\
Frappe & CF & FinalNet & 6, 6 & 0.300 & 1.303 & 0.300 & 4.3 & -- \\
ML-1M & CF & QNN-$\alpha$ (four heads) & 6 & 0.300 & 3.008 & 0.693 & 4.3 & 6/6 \\
 & KD & QNN-$\alpha$ (four heads) & 3 & 0.300 & 3.867 & 0.629 & 6.1 & 3/3 \\
\bottomrule\end{tabular}
\end{table}

\subsection{Towers That Differ Only in Data Order}\label{app:order-only}
On Criteo we trained a second tower from the initialization of a reference tower, changing only the random order in which training examples are drawn, and compared this pair with a pair of towers from two different seeds. All four towers have the structure of Table~\ref{tab:app-arch} but were trained under the FuxiCTR rule of Appendix~\ref{app:train}, an earlier protocol than that of our main Criteo runs, so their AUC is not comparable with Table~\ref{tab:exp-main}. On the validation set, the logits of the order-only pair disagree 95\% as much as those of the seed pair, measured as half the root-mean-square logit difference (0.116; seed pair 0.122). Pooling the order-only pair in logit space brings the Logloss $7.0\times10^{-4}$ below the mean of its members (seed pair $7.8\times10^{-4}$) and raises the AUC over the reference tower by $6.3\times10^{-4}$ (seed pair $7.1\times10^{-4}$). The two towers nevertheless reach different solutions: at the midpoint of the straight line between their weights, the AUC is $1.2\times10^{-2}$ below that of the weaker endpoint (seed pair $4.9\times10^{-2}$). This is a single pair from one initialization on one benchmark, and we did not distill a student from towers that differ only in data order.

\paragraph{Relation to ensemble distillation for CTR prediction.} These results bear on how the gain of ensemble distillation in CTR prediction has been read. In~\cite{ensemblekd2020}, the teachers differ in architecture or in the split of the training data, and the advantage of three teachers over one is attributed to the diversity of different teachers. Their six-teacher ensemble of different architectures adds to each of the three architectures a second copy retrained from a new random initialization, and the student improves further; the authors report this as a gain from applying more teachers. The number of teachers, however, does not by itself produce a gain. By \eqref{eq:teacher-gain}, pooling lowers the Logloss below that of the average member by exactly the disagreement $A_{\rm seed}(K)$ among the members' logits, and an exact copy of a member leaves both the pooled logit and the average member unchanged. A retrained copy helps because its training took a different path through the same data to a different solution, as the order-only pair above does. That work does not single out this source: none of its ensembles varies the seed or the data order alone.

\bibliographystyle{ACM-Reference-Format}
\bibliography{refs}

\end{document}